\documentclass[review]{elsarticle}

\usepackage{lineno,hyperref,amsmath,tabularx}
\modulolinenumbers[5]

\journal{Astronomy and Computing}

\biboptions{authoryear}

\begin{document}

\begin{frontmatter}

\title{Quasi-Monte Carlo Radiative Transfer}

\author{S. G. Shulman\corref{mycorrespondingauthor}}
\cortext[mycorrespondingauthor]{Corresponding author}
\ead{sgshulman@gmail.com}

\address{Lunate PTE. LTD., 336 Smith Street \#05--301, New Bridge Centre, Singapore 050336}

\begin{abstract}
    We consider an alternative to the Monte Carlo method for dust continuous radiative transfer simulations: the Quasi-Monte Carlo method.
    We briefly discuss what it is, its history, and possible implementations.
    We compare the Monte Carlo method with four pseudo-random number generators and five Quasi-Monte Carlo implementations
    using different low-discrepancy sequences and the Hammersley set.
    For the comparison, we study different test matter geometries and problems.
    We present comparison results for single scatterings of radiation from a point source,
    multiple scatterings of radiation from a point source, and single scatterings of radiation from a spherical star.
    In all cases, Quasi-Monte Carlo shows better convergence than Monte Carlo.
    In several test cases, the gain in computation time to achieve a fixed error value reached 40 times.
    We obtained ten times speed up in many of the considered tests.
\end{abstract}


\end{frontmatter}


\section{Introduction}

We obtain most information about stars and interstellar matter by analyzing their radiation.
Many objects have a very complex nature, so we learn about them by analyzing of many radiation characteristics.
Many studies are devoted to modeling how radiation interacts with matter.
They investigate integral quantities (e.g., luminosity and degree of polarization), SEDs, or spectral line profiles.
With the advent of large telescopes that make it possible to obtain images of objects,
many works are devoted to modeling images of stars and the circumstellar matter near them.
Below, we examine dust continuum radiation transfer in a specific spectral band and modeling images.
But it is worth remembering that the world of radiation transfer is much more diverse.

It is important to note that radiative transfer modeling in the circumstellar matter is a complex task.
In the simplest case, to take into account single scatterings from a point source of radiation,
we need to calculate for each point in space how much radiant energy will reach it from the radiation source,
scatter toward the observer, and reach the observer from this point.
It immediately gives us a threefold (volumetric) integral,
even though we omit the calculation of optical depths needed to determine what fraction of the radiation will reach from the star to the scattering point and from the scattering point to the observer.

Actually, taking into account only single scatterings may not be enough.
The contribution of multiple scatterings can be significant for all studied radiation characteristics.
From a mathematical point of view, this means increasing the dimension of the integral.

Some tasks can be simplified using symmetries or some other assumptions.
Then, in rare cases, the analytical solution exists.
However, in most cases, the integral has to be calculated numerically.

Various numerical methods were used to calculate the integral.
We will mention just a few of them:
variable Eddington tensor method \citep{Dullemond2002},
second-order finite differences method \citet{Steinacker2003},
and fifth order Runge–Kutta integration \citet{Steinacker2006}.
\citet{Steinacker2013} review describe the variety of radiative transfer problems and the methods used to solve them.

The Monte Carlo method, based on the simulation of random photon beams, has become a popular method for modeling radiative transfer.
During its use, the Monte Carlo method received a number of important improvements~\citep[e.g.][]{YusefZadeh1984}, modifications and applications.
The Monte Carlo method is applied to many problems of radiative transfer.
For example, the dust continuous radiative transfer \citep{Gerin2017, Muthumariappan2017, Sallum2017},
spectra modelling \citep{Gillanders2022, Fisak2023},
or ionization studies \citep{Harries2017, Vandenbroucke2018, Petkova2021}.
We suggest the reader familiarize himself with the following reviews on radiative transfer by~\citet{Whitney2011}, \citet{Steinacker2013}, and \citet{Noebauer2019}.
It is worth noting that the Monte Carlo method has four big dignities:
it has a proven convergence;
it is easy to implement;
if you get a result with poor quality, you can easily continue the simulation;
it can be easily parallelized.
All these have also led to the emergence of good computation programs with rich capabilities,
which makes its use even simpler and more convenient.
Some of them are \textsc{mcfost} written by~\cite{Pinte2006},
\textsc{hyperion} by~\cite{Robitaille2011},
\textsc{radmc-3d} by~\cite{Dullemond2012},
\textsc{ho-chunk} by~\cite{Whitney2013},
\textsc{torus} by~\cite{Harries2019},
and \textsc{lemon} by~\cite{Xiaolin2021}.

I have long been interested in why the Monte Carlo method is so reasonable and widespread for radiative transfer
and whether there is anything better than simulating random photon packets.
I tried to build a solution based on a discrete grid of directions~---\textsc{DGEM} \citep{Shulman2018} and obtained a more complex compared to Monte Carlo solution.
From a mathematical point of view, \textsc{DGEM} uses Riemann sums to calculate the integral.
This solution is much more effective for simulating single scatterings from a point source.
In the case of multiple scatterings, the solution began to lose to the Monte Carlo method.
The best choice was to consider only the first scattering of the photon packet with DGEM and use the Monte Carlo method for the subsequent ones.
Since then, I improved the method but have not succeeded in adapting it for multiple scatterings or even single scatterings of radiation from a spherical source.

As a result, I focused on literature research in related fields and found very successful examples of using Quasi-Monte Carlo.
It is used in different areas.
Among them, computer graphics should be highlighted, where it is applied to various things, including ray tracing.
As an example, it is worth mentioning \textsc{RenderMan} renderer of \textsc{PIXAR}~\citep{Christensen2018}.
Taking successful applications of Quasi-Monte Carlo in related fields into account,
I thought it would be promising to try it for our radiative transfer simulation, too.
In subsection~\ref{subsec:quasi_monte_carlo}, we briefly describe its history and the main idea behind the Quasi-Monte Carlo method.
Then, in subsection~\ref{subsec:error_estimation}, we present the results of the mathematical justification of why the Quasi-Monte Carlo method should work better than the Monte Carlo method.
It also explains why they are better than classical integration methods, such as Riemann sums, in certain circumstances.

The rest of the paper compares Quasi-Monte Carlo with Monte Carlo for dust continuum radiative transfer in a number of test cases.
Section~\ref{sec:basic_physics} provides a short physics description.
Section~\ref{sec:implementation_details} provides some implementation details of the code and a list of used pseudo- and quasi-random number generators.
Section~\ref{sec:comparison} introduces norms to analyze the solution accuracy.
Section~\ref{sec:point_source_single_sc} is devoted to analyzing the method efficiency for single scattering simulation of radiation from a point source.
Section~\ref{sec:point_sourse_ms} is for multiple scattering modeling.
In Section~\ref{sec:single_scattering_sphere_source}, we perform testing for a sphere source of radiation.
Finally, in Section~\ref{sec:conclusion}, we discuss our results and provide some recommendations for using the Quasi-Monte Carlo method.

\subsection{Quasi-Monte Carlo method}\label{subsec:quasi_monte_carlo}

The classical Monte Carlo method involves the use of random numbers.
In practice, it is possible to use hardware random generators~--- HRNG~\citep{Turan2018, Saarinen2020}.
They are also known as true random number generators (TRNG) or physical random number generators.
However, this approach has some difficulties and limitations:
we have to control physical source entropy and need time to collect enough noise for subsequent random numbers.
Hence, people commonly use pseudorandom number generators (PRNG) for Monte Carlo simulations.
These generators use deterministic algorithms to produce a sequence of numbers with the same properties as true random numbers.
There are different pseudorandom number generators.
Below, we use four of them.

The practice shows that randomly generated points are not very evenly distributed in space.
When we use them to evaluate an integral, it seems intuitive to want a more even point distribution in space.
As a result, beyond the Quasi-Monte Carlo method lies the idea of using a point sequence with a more even space distribution than in the case of random ones.
It is good to use a low-discrepancy sequence for this purpose.

A low-discrepancy sequence is an infinity sequence $(x_1, x_2, x_3, \dots )$ where for all values of $N$,
its subsequence $(x_1, x_2, \dots, x_N)$ has a low discrepancy.
The concept of \textquoteleft discrepancy\textquoteright\ can be introduced in various formal ways depending on the problem.
For our purposes, it is enough to limit to a greatly simplified idea:
the sequence discrepancy is low if the proportion of its points falling into an arbitrary subset $B$ of the domain is close to proportional to the measure of $B$.
In the case of a domain dimension $s$ greater than 1, the low-discrepancy sequence takes $s$ into account.
So, the sequence consists not of numbers but of the $s-$dimension vectors.
As a result, it gives a sequence of quasi-random points that cover the domain of interest significantly faster and evenly compared to random ones.

The history of the low-discrepancy sequence starts from the one-dimensional \citet{vanderCorput1935} sequence.
For a detailed historical review, we recommend~\citet{Faure2015}\@.

In this paper, in the Quasi-Monte Carlo method, we use four well-known low-discrepancy sequences suggested by
\citet{Halton1960}, \cite{Sobol1967}, \citet{Faure1981}, and \cite{Niederreiter1988}.
In addition to infinite point sequences, we also use one finite set proposed by~\citet{Hammersley1960}.

We hope the general idea behind Quasi-Monte Carlo is clear enough.
In subsection~\ref{subsec:generators}, we provide some details about the generators we use.
For our purposes, it is not necessary to discuss the technical details of constructing sequences and their implementations.
For interested readers, the original papers may be a great choice.

\subsection{Error estimations}\label{subsec:error_estimation}

Above, we have a general discussion of the reasons why Quasi-Monte Carlo may be a good choice for radiative transfer simulations.
Now it is time for mathematical error estimation for Monte Carlo, Quasi-Monte Carlo, and classical Riemann sum integration,
which was used in \textsc{DGEM}~\citep{Shulman2018}.

\subsubsection{Error estimation: Monte Carlo}

The error estimation of the Monte Carlo method has been discussed in detail in many papers
(\citealp{Wood1996, Gordon2001, Whitney2011, Steinacker2013}).
We will not detail the derivation and confine ourselves to the main result.
In the simplest case of the unweighted Monte Carlo method,
the error in intensity is the Poisson statistical error $1 / \sqrt{N}$ where $N$ is the number of photon packets.

We use the weighted Monte Carlo method.
In this case, the error estimation is more complicated and depends on the dispersion in the average properties of the photon packets.
However, the matter distribution determines this dispersion.
Hence, the method asymptote remains $1 / \sqrt{N}$.
The proportionality coefficient of error and $1 / \sqrt{N}$ depends on the model.
In the case of the polarized radiative transfer, errors of other Stokes parameters are also estimated in this way.

Thus, in the Monte Carlo method, the error decreases as $1 / \sqrt{N}$
with a coefficient depending on the method implementation and the considered problem.

\subsubsection{Error estimation: Quasi-Monte Carlo}

\cite{Morokoff1994} estimated the errors for the integration with low-discrepancy sequences.
We will not repeat the derivation and confine ourselves to the result.
The errors decrease as $O(N^{-0.5})$ at small $N$, which is the same order as in the Monte Carlo method,
and as $O((\log N)^s / N)$ at large $N$.
The transition value of $N$, at which the asymptote changes, grows exponentially with dimension $s$.
The constant hidden by the big $O$ notation may also be estimated, but we offer it redundant here.

\cite{Faure2010} also perform error estimation for low-discrepancy sequences and get similar results
and note the need to be careful with the asymptotic error estimations.
Even for relatively large values of $N$, we may not yet enter the asymptotic mode and obtain errors very different from the estimations.

These asymptotic error estimations allow us to expect the Quasi-Monte Carlo method to be much faster than the Monte Carlo method.
At least it should not be slower compared to it.
Below, we test this assumption in practice for different models.

\subsubsection{Error estimation: Riemann sum}

Using the Riemann sum for integration, we split the studied volume into sub-volumes and take a point inside each sub-volume.
We summarize function values at these points.
For a good result, all sub-volumes should not be too elongated.
When we increase the number of sub-volumes, their sizes should decrease.

An error estimation depends on the sub-volume sizes and the integration point selection inside the sub-volume.
When all sub-volumes have the same shape and size $r^s$, where $s$ is the space dimension,
we can easily achieve the convergence of $O(N^{-1/s})$.
With a good choice of points (e.g., in the middle of the sub-volume), we can obtain better asymptotic behavior: $O(N^{-2/s})$.

In 1-dimensional space, the errors of the Riemann sum integration with different point choices are $O(N^{-1})$ and $O(N^{-2})$.
Obviously, both ways are much better than $O(1 / \sqrt{N})$ for the Monte Carlo method.
For 3-dimensional radiative transfer problems, the dimension is at least 3
(when we consider a sphere source of radiation or multiple scatterings, the dimension increases).
Hence, the best asymptotic behavior for such a problem is $O(N^{-2/3})$,
which is still better than in the case of Monte Carlo.

For $s$ greater than 4, with a good points choice, e.g., trapezoidal rule, the error asymptotic is $O(N^{-2/s})$.
Its behavior is not better than in the Monte Carlo method.
This phenomenon is named the \textquotedblleft curse of dimensionality\textquotedblright~\cite[see, e.g.,][]{Niederreiter1992}.

\cite{Shulman2018} used Riemann sum integration for the 3-dimensional radiative transfer simulations.
The obtained solution was more complicated than the Monte Carlo solution,
but demonstrated good results for single scatterings from a point source of radiation.
It was less effective than Monte Carlo for multiple scattering modeling,
so we used Riemann sums only for the first scatterings and Monte Carlo for later scatterings.
This combination worked rather well but had few perspectives for effective implementation for a sphere source.

\subsubsection{Error estimation results}

The error asymptotic behavior estimation leads to a conclusion that for problems with dimensions above 4, it is better to use Monte Carlo instead of Riemann sums.
It is consistent with our previous result that DGEM for the six-dimensional problem (double scattering from a point source) no longer gives any benefit.
In our recent experiments, we also did not manage to get any benefit from DGEM in the five-dimensional case (single scattering from a sphere source).
The Quasi-Monte Carlo method should always be not worse than the Monte Carlo.
Moreover, from a certain point, it should become better.

Nevertheless, for multiple scatterings, the problem becomes more complicated.
Accordingly, the more significant the contribution of multiple scattering,
the more complex the situation is.
Hence, the expected gain from using Quasi-Monte Carlo decreases.

We test the latest theoretical results in practice below.

\section{Basis physics}\label{sec:basic_physics}

In this paper, we omit a detailed description of the radiative transfer physics.
We rely on precisely the same physical approaches as in our previous study~\cite{Shulman2018}
where a more detailed description is presented.
One can find good explanations of the radiative transfer physics and computation in
\cite{Chandrasekhar1960} and various reviews, e.g., \cite{Whitney2011, Steinacker2013, Peest2017}.
Below, we briefly discuss the fundamental things in terms of their implementation in the code.
In this study, we consider sphere dust with constant optical properties.

\subsection{Optical depth and albedo}

The dust density $\rho$ distribution over the studied area may differ in various tests.
It is described below for each test.
Based on the extinction opacity, $\kappa$, we obtain the optical depth integrating $\rho \cdot \kappa$ along the line.
Here, we use a Cartesian grid to represent the density distribution for computation algorithms comparison.
The code also supports an unstructured tetrahedral grid as well.
This grid may have smaller tetrahedrons in areas with large density gradients. So, it may describe complicated distributions with high accuracy.
For the tests below, this more complicated approach is not required.

The albedo $\omega \in [0, 1]$ is the ratio of the scattered flux relative to the extinct one.
In our implementation, we weight the photon packet at each scattering by the albedo.

\subsection{Polarization}\label{subsec:polarization}

We study polarized radiative transfer.
To describe the polarization, we use Stokes vector $\mathbf{S} = [I, Q, U, V]$.
$I$ is the intensity,
$Q$ and $U$ are linear polarizations with different alignments relative to some observer's axis,
and $V$ is the circular polarization.

Stokes vector $Q$ and $U$ components depend on the reference frame.
We have to take it into account and rotate reference frames in a proper way when we compute scatterings.
Stokes vector changes during a scattering to a specific direction may be described as the multiplication on the M\"{u}ller matrix
\begin{equation}
    \mathbf{R(\Theta)} = \begin{bmatrix}P_{11} & P_{12} & P_{13} & P_{14} \\
    P_{21} & P_{22} & P_{23} & P_{24} \\
    P_{31} & P_{32} & P_{33} & P_{34} \\
    P_{41} & P_{42} & P_{43} & P_{44} \end{bmatrix}.
\end{equation}
Here $\Theta$ is the scattering angle.
As we study sphere dust, the matrix has only four different nonzero elements:
$P_{11}=P_{22}$, $P_{12}=P_{21}$, $P_{33}=P_{44}$, and $P_{34}=-P_{43}$.
The phase function is used to select a new direction.
Although the code may work with tabular values for M\"{u}ller matrix elements and phase function,
in the current paper, for simplicity, we use an analytical approximation for all these parameters.
The approximations are~\cite{Henyey1941} phase function with the scattering asymmetry parameter $|g| < 1$
\begin{equation}
    x_{HG}(\Theta)=\frac{1-g^2}{\left(1+g^2-2g\cos\Theta \right)^{3/2}}
\end{equation}
and~\cite{White1979} approximation for polarization:
\begin{equation}\begin{split}
    P_{11}=P_{22}&=\left(1-g^2\right)/\left(1-2g\cos\Theta+g^2\right)^{3/2} \\
    P_{12}=P_{21}&=-p_l P_{11}\left(1-\cos^2\Theta\right)/\left(1+\cos^2\Theta\right) \\
    P_{33}=P_{44}&=2\cdot P_{11}\cos\Theta / \left(1+\cos^2\Theta\right) \\
    P_{43}=-P_{34}&=-p_c P_{11}\left(1-\cos^2\Theta_f\right)/\left(1+\cos^2\Theta_f\right),\\
    \Theta_f&=\Theta\left( 1 + 3.13\cdot e^{(-7\Theta/\pi)}\right).
\end{split}\end{equation}
In these equations $p_l$ is the maximum linear polarization,
and $p_c$ is the peak circular polarization.

\subsection{Dust properties}\label{subsec:dust_properties}

In this paper, we use the following dust parameters from~\cite{Shulman2018} for all test cases:
the albedo $\omega = 0.4$,
the scattering asymmetry parameter $g = 0.41$,
the maximum linear polarization $p_l = 0.51$,
the peak circular polarization $p_c = 0$,
and $\kappa = 66$ cm$^2$ g$^{-1}$.
Here, we assume that the gas-to-dust ratio in all test cases is considered the same.
It is equal to 100:1.
These dust properties are purely for testing.
Their selection is rather conventional.
In our previous study, we concluded that the difference in the methods effectiveness slightly depends on the specific dust parameters.
Only an increase in the albedo makes multiple scattering more significant,
which affects the effectiveness of the methods in the multiple scatterings case and reduces the possible speed up.

\section{Implementation of numeric methods}\label{sec:implementation_details}

In the current paper, we consider two different radiative transfer methods:
Monte Carlo radiative transfer and Quasi-Monte Carlo radiative transfer.
These two approaches are very close, so we discuss in detail common parts of the implementation
and describe some nuances when using the Quasi-Monte Carlo method.

Both methods are based on the photon packet propagation through the grid.
\cite{Whitney2011} described the method in detail.
Here, we provide a brief technical explanation.
Pseudo- or quasi-random numbers determine the propagation direction and optical depth.
The difference is how we obtain these numbers: we discuss generators in subsection~\ref{subsec:generators}.
So, the source emits the photon packet in some direction (\ref{subsec:photon_from_point_source} and \ref{subsec:photon_from_sphere_source}),
it propagates through the matter (\ref{subsec:photon_propagation}),
scatters (\ref{subsec:photon_scattering}), and finally the observer registers the photon packet~(\ref{subsec:producing_images}).
Due to the slight technical difference between method implementations, we have a general code for optical depth integration, Stokes vector changes, and photon packet registration.
In all cases, \textquoteleft the photon packet modeling\textquoteright\ means the way to obtain scattered light.
It is enough to model only one photon packet propagating from the source toward the observer to obtain direct light from the point source.
It requires only one optical depth integration, so it does not contribute significantly to the total computation time.
For the sphere source, it is more complicated, but in our tests, we use a \textquoteleft moon\textquoteright\
to exclude the star from the result.
Hence, we work only with the scattered light.

Profiling shows that, in our tests, the most computationally expensive function is the optical depth integration.
Using one implementation for all methods makes the comparison more resistant to minor optimization inside the methods.

In this paper, we use a Cartesian coordinate system $xyz$ and spherical coordinates $(\phi, \theta)$.
The azimuthal angle $\phi$ is an angle in $xy$ plane from the $x$ axis positive direction
and the polar angle $\theta$ is measured between from $z$ axis positive direction.

\subsection{Random number generators}\label{subsec:generators}

Random number generator provides numbers with uniform distribution from 0 to 1.
Hence, there are two parameters in the method implementation: generator (with seed for Monte Carlo method) and a total number of photon packets.
Increasing the second one leads to better result quality.
We use a weighted Monte Carlo method where photon packets have computational weights initially equal to 1.
If we have several light sources, the photon number should be divided between them proportionally to the luminosities.

To compare Monte Carlo and Quasi-Monte Carlo radiative transfer approaches, we use four pseudo-random generators and four quasi-random generators.
All these generators are described in the literature. Their implementations are available in different sources (including our repository).
Hence, we omit their implementations and explanations here.

The pseudo-random number generators are:
\begin{itemize}
    \item Minimum Standard \citep{Park1988, Park1993}.
    \item \cite{LEcuyer1988}.
    \item Ranlux48 \citep{Luscher1994} see also \citep{James1994}.
    \item Mersenne Twister \citep{Matsumoto2000}.
\end{itemize}

The quasi-random number generators are:
\begin{itemize}
    \item \cite{Sobol1967}. Our implementation is based on \cite{Bratley1988} and \citet{Joe2008} studies.
    \item Niederreiter base 2 \citep{Niederreiter1988}. Implemented following \cite{Bratley1992}.
    \item \cite{Faure1981}.
    \item \cite{Halton1960, Halton1964}.
\end{itemize}

These quasi-random number generators are for infinite number sequences.
In practice, we use some specified number of photon packets.
This information is suitable for creating a finite set of points with even better space distribution.
A typical example is the \citet{Hammersley1960} set.
The $s$-dimension Hammersley set uses $s-1$ dimension Halton sequence and for the last dimension uses values
$\tfrac{i}{N}$ where $i$ is from $1$ to $N$.
$\tfrac{i}{N}$ gives very good discrepancy for that dimension.
In this paper, we also use the Hammersey set, but we use $\tfrac{i}{N}$ sequence for the first dimension instead of the last.

For Minimum Standard, Ranlux48, and Mersenne Twister generators, we use C++ Standard Template Library implementations.
For other generators, there are custom implementations based on the papers.
L'Ecquyer generator implementation is inherited from MCPOLAR~\cite{MCPOLAR}.
Some useful notes about quasi-random generations and their efficiency are also presented in \cite{Fox1986}.
For all quasi-random generators, leading zeros are omitted.
We do not omit nonzero values, which may be recommended in some cases for better behavior at the beginning of the sequence.
In our usual cases with a huge number of photon packets, it seems not so important.

Different pseudo- and quasi-random number generators take different times to generate a number.
In our computation problems, the optical depth integration through the grid may consume more than 95\% of the time.
Random number generating takes a tiny part of the program's running time.
We can neglect it.

Another question concerns memory consumption.
Memory consumption for pseudo-random number generators is constant and insignificant for modern computers.
For quasi-random number generators, memory consumption depends linearly on the problem dimension.
Typically, the generator requires some constant memory and 8 or 16 bytes for each dimension.
With a reasonable dimension of the problem, we are talking about tens or hundreds of bytes for the generator.
It can also be neglected.

Hence, memory consumption and performance of the random number generator should not be considered reasonable for choosing a specific generator.

\subsection{Photon packets from a point source}\label{subsec:photon_from_point_source}

Photon packets from the point source are isotropically distributed in the space.
We describe the propagation direction with spherical coordinates where
$\theta \in [0, \pi]$ is a polar angle and $\phi \in [0, 2\pi]$ is an azimuthal angle.
We need two random numbers $u$ and $v$ to obtain a direction for a new photon packet
\begin{equation}\begin{split}\label{eq:point_source_propagation_direction}
    \phi &= 2\pi \cdot u, \\
    \theta &= \arccos(2v - 1). \\
\end{split}\end{equation}

The last thing we should determine for the photon packet from the star is the position of scattering.
We compute the optical depth $\tau_1$ along the propagation direction until the grid edge.
If $\tau_1$ is close to zero (less than the cut-off value $\tau_{\min}$), we finish this photon packet modeling.
Otherwise, we compute the weight of the scattered photon packet (the fraction of scattered radiation), which is $1-\exp{\left(-\tau_1\right)}$.
After that with an additional random number $w$ we obtain new optical depth
\begin{equation}\label{eq:random_depth}
    \tau = -\ln\left[1 - w (1 - \exp{\left(-\tau_1\right)})\right].
\end{equation}
The photon packet propagates through the matter with the optical depth $\tau$ and scatters.
The optical depth integration along the propagation direction gives us the scattering position.

\subsection{Photon packets from a sphere source}\label{subsec:photon_from_sphere_source}

Modeling photon packets from a sphere source is a bit more challenging.
In addition to the photon packet propagation direction and optical depth, we have to determine the point of photon packet emission on the surface of the sphere source.
It increases the problem dimension by 2.
For simplicity, we do not take into account the limb darkening.
We assume that the radiation from the sphere source is isotropic.

So, we need five random numbers per one photon packet.
Typically (like in HYPERION source code \citet{Robitaille2011}), we use two random numbers $v_1$ and $v_2$ to obtain the direction to the emission point from the center of the sphere source in spherical coordinates $(\phi_e, \theta_e)$:
\begin{equation}\begin{split}\label{eq:sphere_source_emit_point}
    \phi_e &= 2\pi \cdot v_1, \\
    \theta_e &= \arccos(2v_2 - 1). \\
\end{split}\end{equation}
Taking into account the known radius of the source, this gives us an unambiguous position in space.

After this, we use two more random numbers $v_3$ and $v_4$ to determine the propagation direction of the photon packet in the local coordinate system:
\begin{equation}\begin{split}
    \phi_l &= 2\pi \cdot v_3, \\
    \theta_l &= \sqrt{v_4}. \\
\end{split}\end{equation}
By rotating local angle $(\phi_l, \theta_l)$ on the position angle of the emission point $(\phi_e, \theta_e)$, we obtain the propagation direction in the global coordinate system.
The last, fifth random number $v_5$ characterizes the optical depth in the same way as for the point source in~\ref{eq:random_depth}.

The described approach works fine for Monte Carlo simulations.
But in Quasi-Monte Carlo simulations, there are reasons to try another approach.
We can expect that the number distribution in later dimensions is worse compared to early dimensions.
In this case, we may want to use random numbers with the worst distribution for the emission point position on the star surface.
The idea is straightforward: if the star is small enough to treat it as a point source (but we are not sure about it), we want
the computation of photon packets from the sphere star to be as good as the computation of the radiation from a point source.

To implement this behavior, we use another random numbers order:
$v_1$ and $v_2$ are used to determine the propagation direction as $u$ and $v$ in~\ref{eq:point_source_propagation_direction} for the point source.
$v_3$ is used as $w$ from~\ref{eq:random_depth}.
$v_4$ and $v_5$ are for the direction to the emission point from the center relative to the propagation direction (in a local coordinate system).
This relative direction is obtained from the same equation as~\ref{eq:sphere_source_emit_point}.
After that, we rotate the relative direction to the emission point on the position angle of the propagation direction
to obtain the direction to the emission point in the global coordinate system.
It allows us to compute the emission point coordinates.

The comparison of these two approaches --- straightforward and inverse order --- is discussed in the later sections.

\subsection{Photon packet propagation}\label{subsec:photon_propagation}

For a more accurate comparison of different methods,
it is principal that the differences are associated only with the features of photon emission and scattering point determination.
We have implemented all methods within one program.
Therefore, many functions are used for all methods.
It eliminates unnecessary differences between method implementations.
Most of the remaining functions are fairly simple and usual for radiative transfer, so we will not dwell on them for a long time.
Below, we list the main general things and briefly describe the integration of the optical depth.

All changes of the Stokes parameters are implemented once for all methods according to physical and geometry properties from Sec.~\ref{subsec:polarization}.
We identically take into account the albedo, multiplying the weight of the photon packet by it.
We always use the peel-off technique described in Sec.~\ref{subsec:producing_images} to bin photons on the image.

The most computationally expensive thing is optical depth integration.
For integration speed up, a density mesh is used.
We use a Cartesian grid to approximate density when comparing methods
(it is described in detail by~\citealp{Whitney2011}).
In each cell of this mesh, the matter density $\rho$ is assumed to be constant.
The extinction opacity $\kappa$ is assumed to be constant for the entire mesh.

We start integration from the cell with a photon packet.
We use photon packet position to determine the next cell along the ray and distance $d_{cell}$ to the cell edge.
So, optical depth in this cell is $\tau_{cell} = d_{cell} \kappa \rho$.
The optical depth along the ray to the mesh edge is the sum of all $\tau_{cell}$ which are crossed by the ray.

When we move a photon packet on the specified optical depth $\tau$,
we go along the propagation direction accumulating $\tau_{cell}$.
When accumulated optical depth exceeds $\tau$,
we stop the process and calculate the final position of the photon inside the cell based on the $\tau_{cell}$,
the required optical depth and the length of the ray segment inside this cell.

In our code, we made several low-level optimizations to speed up the Cartesian grid integration.
The code is published, so we will not describe these optimizations here.

For actual calculations, other more flexible grids can be used.
We use density distribution equations to precompute matter density in the mesh cells.
They are described in the following sections for each test.

In the code, direct light is taken into account separately. For this purpose, we integrate optical depth from the source toward the observer.
In the tests, we usually compare only scattered radiation. The direct light is the same.
Hence, there is no sense in comparing it.

\subsection{Photon packet scattering}\label{subsec:photon_scattering}

We start photon packet scattering by creating an additional photon packet scattered in the observer's direction and projecting it into an image.
This technique is well-known as \textquotedblleft peel-off\textquotedblright~\cite{YusefZadeh1984}.
\cite{Pozdnyakov1977} used before a close approach and added a weighted photon packet to a spectrum after each scattering.
The placing of photon packets into the image is described in Sec.~\ref{subsec:producing_images}.

After peeling-of we take \citealp{Henyey1941} phase function with the scattering asymmetry parameter $g$ into account.
We use two random numbers $u$ and $v$ to compute scattering angle $\Theta$ and additional orientation angle $i_1$:
\begin{equation}\begin{split}
    \cos \Theta&=\frac{1+g^2-\left[\left( 1-g^2 \right)/\left( 1-g+2gv\right)\right]^2}{ 2g }, \\
    i_1 &=2\pi\cdot u.\\
\end{split}\end{equation}
These two angles unambiguously determine the new direction of photon packet propagation.
(see \citealp{Chandrasekhar1960, Shulman2018} for detailed scattering geometry).

Before the next scattering, the photon packet propagates through the optical depth $\tau_2 = - \ln w$,
where $w$ is also a random number.
The scattering position is determined by the optical depth integration.
During the integration, the photon packet may leave the grid. In this case, we finish its modeling.
In general, the photon packet computation finishes when it leaves the grid or after the fixed number of scatterings,
which is the model parameter.

\subsection{Producing images}\label{subsec:producing_images}

We use the peeling-off technique for producing images~\cite{YusefZadeh1984}.
During each photon packet scattering, we produce an additional photon packet,
which is scattered toward the observer and immediately registered on the image.
If there is more than one observer, we produce such a packet for each observer.
The weight of the additional photon packet $W$ depends on the scattered packet weight $W_0$,
an optical depth toward the observer $\tau_{obs}$ and the phase function.
In case of \cite{Henyey1941} phase function we have
\begin{equation*}\begin{split}
    W &= W_0 e^{-\tau_{obs}} \cdot p_{obs}, \\
    p_{obs} &= \frac{1}{4\pi}\frac{1-g^2}{(1+g^2-2g\cos \Theta_{obs})^{3/2}}.
\end{split}\end{equation*}
Here $p_{obs}$ is the probability of the photon scattering toward the observer
with the $\Theta_{obs}$ angle between the propagation direction of the original packet and the direction toward the observer.
In general, $p_{obs}$ may be adopted from other phase functions, including tabulated.
In this work, the Henyey-Greenstein phase function is a reasonable choice for the methods comparison.

After $W$ computation, we just need to add $W$ to the corresponding image pixel (or pixels).
Projection of a point in the 3D world on the 2D image (taking into account the direction toward the observer),
image pixel resolution, pixel density, and possible coronagraph mask are easy.
Below, we describe the direction toward the observer with spherical coordinates $(\phi_{obs}, \theta_{obs})$.

In this paper, the star luminosity ($I_*$) is set as one for all models because it does not affect the intensity distribution throughout the image.
The intensity distributions on all photometric maps are presented in units of the star luminosity.
For converting the presented images to fluxes, we should multiply them by $\tfrac{L}{d^2}$,
where $L$ is the star luminosity and $d$ is the distance to the system.

\subsection{Using Quasi-Random Numbers}\label{subsec:quasi_randoms_usage}

First, we should determine the domain dimension $s$ and get the corresponding generator.
For example, when we model photon packets from point sources, we have a 3D area: two dimensions for propagation direction and one for the optical depth.
If we have a sphere source, we obtain a 5D area: additional two dimensions for the emission position on the sphere.
Every subsequent scattering also gives additional three dimensions for new propagation direction and optical depth.
So, for one photon packet, we use a vector of quasi-random numbers instead of a sequence of independent random numbers.
To avoid high dimensions, we can try to use several quasi-random vectors of lower dimensions.
Thus, to model 4 scatterings from the point source, we can use one 12D vector per photon packet or four 3D vectors per packet (one 3D vector per scattering).
It was not obvious which of these approaches was more suitable for our tasks, so we tried both and compared them in Sec.~\ref{sec:point_sourse_ms}.

Second, we model photon packets in the same way as in the Monte Carlo method but using another random number generator.
Quasi-random number generators may provide for the user consequent numbers instead of vectors.
It means we use elements of the vector one by one and then elements of the next vector.
With such implementation, we have to make only one change in code to use quasi-random numbers instead of random numbers.
If we finish photon packet modeling and use not all numbers from the vector, we should skip the remaining numbers from this vector.
Hence, for every purpose, we should always use numbers from the same position in a vector.

For example, we model 2 scatterings for photon packets from a point source.
We have 6D area: 2 numbers per direction of light propagation from the star, 1 number per optical depth, 2 numbers per new propagation direction after scattering, and the last 1 number per optical depth before the second scattering.
Let us generate propagation direction from the source using 2 numbers and discover that the optical depth in this direction is zero.
We finish modeling this photon packet, but we have to skip four numbers from the generator.
As a result, the next photon packet will use the first two elements of the next vector for propagation direction from the star.
It allows us to use the numbers correctly and calculate effectively.

\section{Comparison of the methods}\label{sec:comparison}

To compare the methods, we use them with different photon packet numbers for several test geometries.
During each test, we measure the computation time and the result quality.
We do not produce various tests with different random number generator seeds as we expect,
that for a huge number of photon packets and many measurements, the seed should not be very significant.

The stability and accuracy of the time measurements are achieved in the following way.
All studied methods are implemented in one program.
For all tests, the program compilation options were the same.
All tests run on one PC with an Intel(R) Core(TM) i5-10400F CPU locked at 4.1 GHz.
During the time measurements, only one core was busy with the computations. All other cores were idle.
These conditions ensure that we do not obtain systematic or random differences between time measurements for different methods.
Even with such limitations, the computation time for the same problem varies from run to run.
Experimentally, we can conclude that the error in time measurements in our tests usually does not exceed 5\%
(without these limitations, it is easy to obtain discrepancies in computing time by 1.5 times on the same PC).
We take this into account while analyzing the results and rely only on more significant time differences.

The program integrates Stokes vectors of all registered photon packets, taking scattering number into account.
It also produces images for different scatterings.
Hence, we are able to compare both integral and image results (pixel by pixel) for different elements of the Stokes vector.
We can compare results for the studied geometry and the contribution of different scatterings.
For example, we can compare the contributions from secondary scatterings.
To estimate the errors of different methods with different parameters, we produce a high-quality Monte Carlo result
for every studied task.
This reference solution uses a separate random seed to avoid unnecessary correlations with the tested Monte Carlo results.

\subsection{Integral result comparison}

The integral light intensities in the reference solution and in the tested solution are $I_{ref}$ and $I$, respectively.
The results may be brighter in a model with one dust distribution than in models with other distributions.
We do not want to take it into account while comparing.
Hence, for the result comparison, we use the relative difference norm:
\begin{equation}\label{integral_norm}
    \delta I = \frac{I - I_{ref}}{I_{ref}}.
\end{equation}
It is a dimensionless value.
Later, we'll usually present such results in percentages.

\subsection{Image comparison}

In many studies, we are interested in the light parameters distribution along the model as well as in integral values.
It is essential both for physical studies and method comparison.
We should know that the suggested method does not provide any noticeable artifacts in some areas.
So, we produce images with absolute and relative differences between images.
We also use an image-based norm for result comparison.
These differences and the norm were previously used in \citet{Shulman2018}.

Let us consider two images $A$ and $B$ with the same size $N_x \times N_y$.
The image-based difference norm is
\begin{equation}\label{norm}
\frac{\Delta I}{I} = \frac{\sum_{x=0}^{N_x-1} \sum_{y=0}^{N_y-1} |A(x,y)-B(x,y)|}{\sum_{x=0}^{N_x-1} \sum_{y=0}^{N_y-1} \frac{|A(x,y)+B(x,y)|}{2}},
\end{equation}
where $x$ and $y$ are pixel coordinates in images.
$A(x,y)$ and $B(x,y)$ are the pixels with corresponding coordinates in the images $A$ and $B$.
The image-based norm obviously depends on $N_x$ and $N_y$ but provides additional information compared to the integral norm.
Here, we use the mean between $A(x, y)$ and $B(x,y)$ in the denominator to avoid zero division.
Later, we also present $\Delta I/I$ in percentages.

\medskip

The norm computations for other Stokes vector parameters are implemented in the same way.
When we have per-pixel differences between images $A$ and $B$, the integral norm is the sum of such differences,
while the image-based norm is the sum of modules of these differences
(we omit the difference between denominators of these expressions).
Hence, the absolute value of the integral norm is usually less than the image-based norm, but the image-based norm is less noisy.

\section{Single scatterings case for a point source}\label{sec:point_source_single_sc}

In this section, we study single scatterings for a point source of radiation.
We use four test cases: a flared disk observed from two directions, a dust sphere envelope,
and a fractal dust cloud.
In~\cite{Shulman2018}, we considered five different observer positions for the flared disk and obtained very close results,
so in the current paper, we limit to two different positions.

In all tests of this section, the star is in the coordinate system origin.
The resulting image size is $200 \times 200$ pixel.
The observed region radius is 800 au.
So, the $1600$~au $\times 1600$~au square is shown.
In images, we present dust matter together with the star,
but we do not take the star into account during the difference norm computations.
All reference images were obtained by the Monte Carlo method with \cite{LEcuyer1988} random number generator
with $5\cdot10^{11}$ photon packets.

\subsection{Test cases}\label{subsec:point_source_single_sc_tests}

\subsubsection{Flared disk}\label{subsubsec:point_source_single_sc_tests_flared}

A flared disk is a widely used geometrical model.
Following other studies (e.g.,~\citealp{Teixeira2009, Robitaille2011}) we describe the flared disk density as
\begin{equation}\label{eq:flared_disk}
\rho(x,y,z) = \begin{cases}
                  \rho_0 \left(\frac{R_0}{r}\right)^\alpha \exp\left[-\frac{1}{2}\left(\frac{z}{h(r)}\right)^2\right],       & R_{i} \le r \le R_d, \\
                  0,& \text{otherwise}.          \end{cases}
\end{equation}

\begin{figure}
    \begin{minipage}[h]{0.495\linewidth}
        \includegraphics[width=\linewidth]{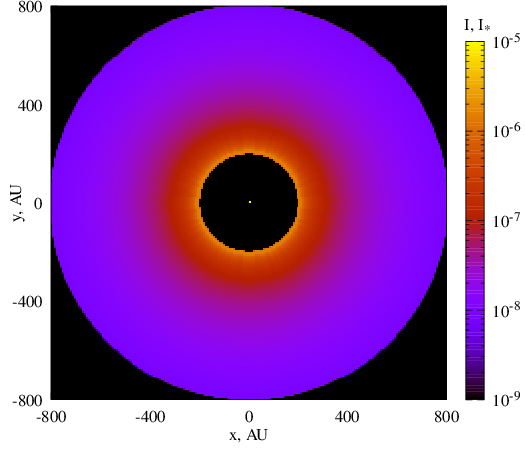}
    \end{minipage}
    \hfill
    \begin{minipage}[h]{0.495\linewidth}
        \includegraphics[width=\linewidth]{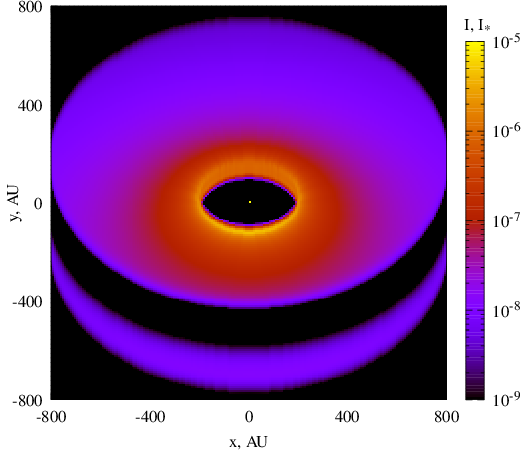}
    \end{minipage}
    \caption{Photometric images of the flared disk observed from the pole (left panel) and with $\theta_{obs} = 45$\textdegree\ (right panel).
        $5\cdot10^{11}$ photon packets were used for the Monte Carlo simulation of both models.}
    \label{fig:ref_flared_disk}
\end{figure}

Here $r = \sqrt{x^2 + y^2}$ is the length of the point projection onto the disk plane ($z = 0$)\@.
$h(r) = h_0\left( r / R_0\right)^{\beta}$ is the disk scale height.

This model has several parameters.
We use the same model parameters as in~\cite{Shulman2018}.
$R_i=200$ au and $R_d=800$ au are inner and outer disk radii,
$h_0 = 7$ au is a disk scale height corresponding to a $R_0 = 100$ au radius,
$\alpha= 2.25$ is the radial density exponent,
$\beta = 1.25$ is the flaring power, and
$\rho_0 = 2.4 \cdot 10^{-15}$ g cm$^{-3}$ is the density normalization constant.
As a result, the disk mass is $\approx 0.027 M_\odot$.

In the first test case, the observer is positioned in a direction with $\phi_{obs} = 0$ and $\theta_{obs} = 0$ (the disk pole).
In the second case $\theta_{obs} = 45$\textdegree\ and $\phi_{obs}$ is the same.
Reference photometric maps for the disk observed from both positions are presented in Fig.~\ref{fig:ref_flared_disk}.

\subsubsection{Dust sphere envelope}\label{subsubsec:point_source_single_sc_tests_sphere}

For the next test, we use a simple sphere envelope model.
The main idea was to have such dust distribution,
which has some scattering matter in every direction from the star.
The envelope is the same as in~\cite{Shulman2018}\@.
It has the following simple shape
\begin{equation}\label{eq:sphere}
\rho(x,y,z) = \begin{cases}
                  \rho_0,       & R_{i} \le r \le R_d, \\
                  0,& \text{otherwise}.          \end{cases}
\end{equation}

Here $r=\sqrt{x^2 + y^2 + z^2}$ is the distance from the star.
$R_i = 500$ au and $R_d = 550$ au are the inner and outer sphere radii,
and $\rho_0=2.4\cdot10^{-17}$ g cm$^{-3}$ is the density.
The envelope mass is $\approx 0.0067 M_\odot$.

\begin{figure}
    \begin{minipage}[h]{0.495\linewidth}
        \includegraphics[width=\linewidth]{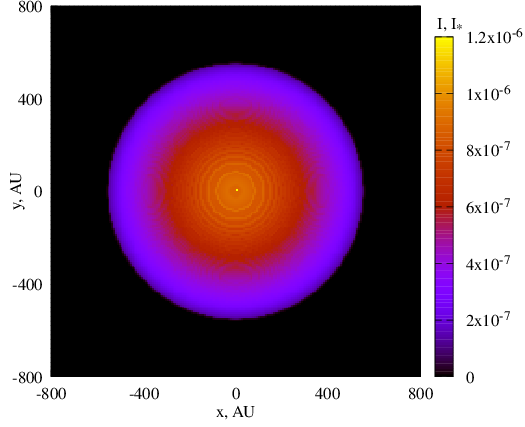}
    \end{minipage}
    \hfill
    \begin{minipage}[h]{0.495\linewidth}
        \includegraphics[width=\linewidth]{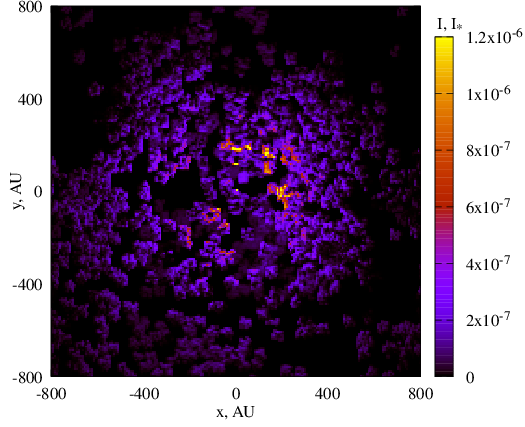}
    \end{minipage}
    \caption{Photometric images of the dust sphere envelope (left panel) and the fractal dust cloud (right panel).
        $5\cdot10^{11}$ photon packets were used for the Monte Carlo simulation of both models.}
    \label{fig:ref_sphere_fractal}
\end{figure}

The observer is positioned in direction with $\phi_{obs} = 0$ and $\theta_{obs} = 0$.
The reference photometric map is presented in the left panel of Fig.~\ref{fig:ref_sphere_fractal}.

\subsubsection{Fractal dust cloud}

Both flared disk and dust sphere envelop provide smooth dust distributions.
For the last test, we use a fractal dust distribution,
which is rather clumpy.

As in the previous work, we follow~\cite{Elmegreen1997, Mathis2002} to obtain hierarchically clumped dust distribution.
The data generation procedure is following
\begin{enumerate}
    \item Consider a cube consisting of cube cells.
          We have $200$ cells along each edge ($200^3$ cells in total).
          Let the cube size be $L_{cube}$.
    \item Place $N$ random points into the cube.
    \item Create a smaller cube for each point.
          The point should be in the center of the cube.
          A smaller cube might be partly outside the large one.
          The size of a smaller cube should be $L_{cube} / (2 \cdot N^{1/D_{cube}})$. $D_{cube}$ is the \textquotedblleft fractal dimension\textquotedblright\ of our distribution.
    \item We place new $N$ random points in every small cube
    \item Repeat steps 2--3 twice more.
          Finally, we have $N^4$ points.
    \item Shift all the points outside the large cube to within it by translating each Cartesian coordinate outside the supercube by $L_{cube}$.
    \item Set the density in every cell of the large cube proportional to the number of points inside it.
    \item Normalize density to achieve desired average radial optical depth equal to $\tau_{cube}$ or obtain specified dust mass.
\end{enumerate}

In our model we have  $L_{cube} = 1600$ au, $N = 32$, $D_{cube} = 2.3$, and $\tau_{cube} = 2$.
For random points generation~\cite{LEcuyer1988} generator with a seed equal to $-1556$ was used.
Successively obtained random numbers triples were used to calculate the coordinates of points.
The observer is positioned in direction with $\phi_{obs} = 0$ and $\theta_{obs} = 60$\textdegree.
The reference photometric map is resented in the right panel of Fig.~\ref{fig:ref_sphere_fractal}.

\subsection{Integral results}\label{subsec:point_source_single_sc_integral_results}

\begin{figure}
    \begin{minipage}[h]{0.495\linewidth}
        \includegraphics[width=\linewidth]{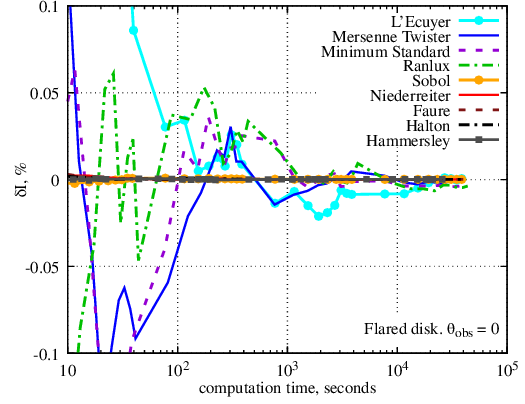}
    \end{minipage}
    \hfill
    \begin{minipage}[h]{0.495\linewidth}
        \includegraphics[width=\linewidth]{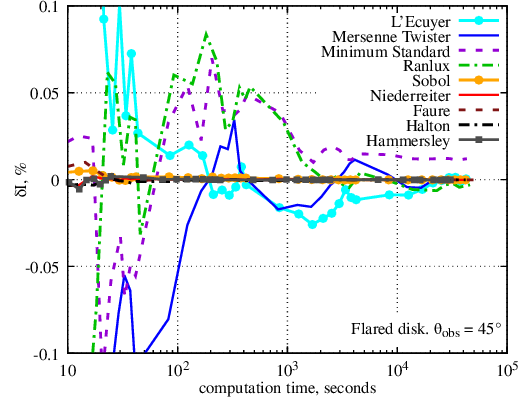}
    \end{minipage}
    \vfill
    \begin{minipage}[h]{0.495\linewidth}
        \includegraphics[width=\linewidth]{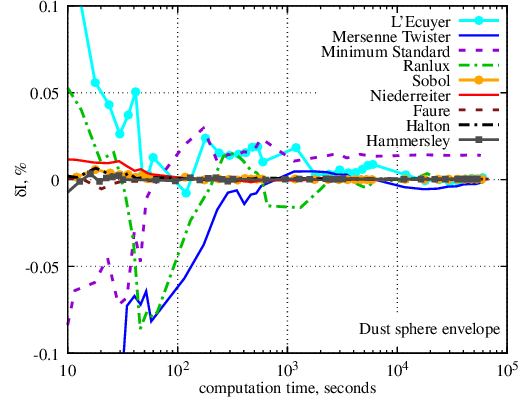}
    \end{minipage}
    \hfill
    \begin{minipage}[h]{0.495\linewidth}
        \includegraphics[width=\linewidth]{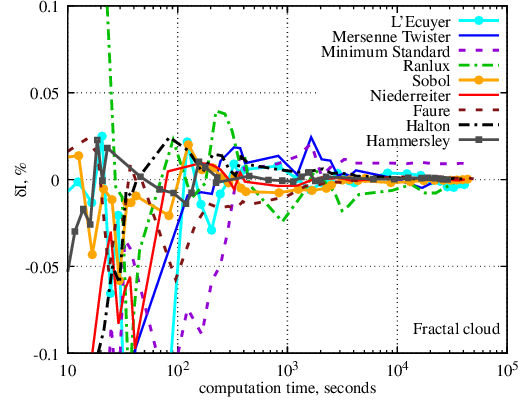}
    \end{minipage}
    \caption{The relative difference norm \ref{integral_norm} for results obtained with different random number generators for
        the flared disk observed from the pole (top left panel),
        flared disk observed with $\theta_{obs} = 45$\textdegree\ (top right panel),
        the dust sphere envelope (bottom left panel), and the fractal dust cloud (bottom right panel).
        The star is considered a point source of radiation. Only single scatterings are simulated.}
    \label{fig:flared_disk_ps_1sc_integral}
\end{figure}

We start the discussion of the results from the integral results.
In Fig.~\ref{fig:flared_disk_ps_1sc_integral}, we present four panels with the results for all tested methods for the four test geometries.

Even a quick analysis of the results shows a number of important features:
\begin{enumerate}
\item The graphs for each method show noticeable fluctuations.
    The error can change the sign and does not decrease monotonously.
\item Quasi-Monte Carlo methods (including the Hammersley set) converge to zero significantly faster (it is evident for the disk and the sphere).
\item Fractal distribution of matter turns out to be the most difficult test, in which $\delta I$ remains significant for all tests even with rather long calculation times of about $10^3$ seconds
    (while for other geometries, Quasi-Monte Carlo solutions have almost zero errors after $10^2$ seconds).
\item The $\delta I$ norm for Minimum Standard generator from a certain moment changes the value in a very narrow range and does not converge to zero.
\end{enumerate}

The fourth feature has a trivial explanation: Minimum Standard generator has a short period $2^{31}-2 \approx 2\cdot 10^9$.
We need $3$ random numbers per photon packet during the single scattering computation.
Moreover, $2^{31}-2$ is a multiple of $3$.
Hence, after the simulation of $\sim 7\cdot10^8$ photon packets, we begin to simulate the photon packets we have already simulated.
Obviously, it does not improve the result.
We observe such behavior in all tests with the Minimum Standard generator in this paper.

The first feature is an obvious consequence of using the Monte Carlo method: we have the sum of random realizations.
So, we can see that the Quasi-Monte Carlo method works better, but it is rather complicated to estimate the exact values.
To estimate typical Monte Carlo method error depending on computation time (or number of photon packets)
we may spend much computation time for simulations with various random number generator seeds.
Unfortunately, the low-discrepancy sequences for the Monte Carlo method are more complicated and often do not provide
us with such an opportunity.
Therefore, to assess the advantage of the Quasi-Monte Carlo method over the Monte Carlo method,
we need some kind of estimation (maybe not very stable) that will characterize the error dependence on time using one sequence of measurements.

As a quite simple approach to this estimation, we can consider approximating the results of numerical simulations.
Unfortunately, this approach has a number of difficulties:
1) at short simulation times, we have a very high noise.
2) at long simulation times, the results for Quasi-Monte Carlo decrease more slowly. Apparently, the error in the reference solution obtained by the Monte Carlo method begins to affect the norm.
Although we obtained reference solutions by simulating $5\cdot 10^{11}$ photon packets, due to the slow convergence of the Monte Carlo method,
we cannot consider them as exact solutions or even as a very accurate one.
3) the Quasi-Monte Carlo convergence changes its asymptotic depending on the number of photon beams.
It makes it impossible to approximate the dependence of error on time with a simple function with good quality.
However, it is worth noting that we used the same set of values for the photon packet number for all generators.
It makes comparison easier than if the points were uneven and distributed differently for different methods.
We use the least-squares method to look at several approaches with different approximations and compare their results.
We omit measurements with a computation time of less than 10 seconds as too noisy and unimportant from a practical point of view.

A simple function to approximate $|\delta I|$ could be $a \times t^{-0.5}$ where $t$ is the computation time.
From theoretical considerations, it should be a good fit for Monte Carlo and a worse one for Quasi-Monte Carlo.
Since the errors are highly unstable and there are many values for all test cases,
we consider an average value for the Monte Carlo method and an average value for classical implementations of the Quasi-Monte Carlo method
(with Sobol, Niederreiter, Faure, and Halton generators).
We have one solution for the Quasi-Monte Carlo method with the Hammersley set. So, the result is presented without averaging.
When the Minimum Standard solution reaches its quality limit, we omit it from the averaging.
The values of $a$ parameter in different approximations are presented in table~\ref{table:integral_norm_sqrt_approximation}.
We can see that the $|\delta I|$ values differ considerably from one test geometry to another.
It depends on the selected method, too.
In our tests, the Quasi-Monte Carlo method gives $|\delta I|$ 3--95 times less than the Monte Carlo method.
Quasi-Monte Carlo with the Hammersley set sometimes gives even better results.
However, the difference between the Hammersley set and quasi-random number generators is insignificant in some cases.
The Hammersley set gives $|\delta I|$ 8--100 times less than the Monte Carlo method.
So, analyzing $1/\sqrt{t}$ asymptotic behavior demonstrates a nice benefit from using Quasi-Monte Carlo.

\begin{table}[htbp]
    \centering
    \begin{tabularx}{\textwidth}{>{\hsize=1.6\hsize} X | >{\hsize=0.8\hsize}X  |>{\hsize=0.8\hsize}X  |>{\hsize=0.8\hsize}X }
        a, \%                           & Monte Carlo & Quasi MC   & Hammersley \\ \hline
        Flared disk. $\theta= 0$             & 46        & $0.5$ & $0.5$ \\
        Flared disk. $\theta= 45$\textdegree & 48        & $0.8$ & $0.7$ \\
        Dust sphere envelope                 & 32        & $1.7$ & $0.9$ \\
        Fractal cloud                        & 61        & $19$   & $7$  \\
    \end{tabularx}
    \label{table:integral_norm_sqrt_approximation}
    \caption{Values of parameter $a$ in percentages for $|\delta I|$ approximation as $a \times t^{-0.5}$ function.
        For each test geometry, the table shows the average values for the Monte Carlo method (L'Equier, Minimum Standard, Mersenne Twister, and Ranlux48 generators),
        the average value for the Quasi-Monte Carlo method (Sobol, Niederreiter, Faure, and Halton generators) and the value for the Quasi-Monte Carlo method with the Hammersley set.}
\end{table}

Another possible $|\delta I|$ approximation may be presented as a bit more complicated function $a \times t^{-b}$.
Here, we have two parameters, which makes the approximation potentially more accurate and less stable.
With such a function, the approximation parameters for the Monte Carlo method results do not change.
In all cases, we obtain $b = 0.5$ with high accuracy, which matches the previous approximation.
For the Quasi-Monte Carlo method (including the version with the Hammersley set),
the situation is more complicated: taking values with short computation times into account, we also get $b \approx 0.5$.
But skipping points with a computation time of less than 100 seconds leads to $b \sim 1$ for most cases (the exception is the fractal cloud test geometry,
where $b$ is still about $0.5$).

More complicated approximations lead to unstable results.
So, we can limit to these two functions.
Taking the instability of the results into account, we should point out only the key features.
We can conclude that the Quasi-Monte Carlo method definitely gives noticeably better result quality than the Monte Carlo method.
It also demonstrates a better convergence rate.
The Quasi-Monte Carlo method with the Hammersley set usually outperforms other considered implementations.

\medskip

We can carry out a similar $a \times t^{-0.5}$ approximation for other Stokes vector parameters.
We do not consider circular polarization. So, it makes sense only to look at $Q$ and $U$ Stokes parameters.
There is no point in such approximation for the $U$ parameter for a flared disk and for a dust sphere envelope.
In these three cases, $U = 0$. Hence, calculating the relative error for it does not make sense.
For parameters $Q$ and $U$, the decrease in the approximation parameter $a$ is sometimes less than for the intensity.
Nevertheless, the Quasi-Monte Carlo method gives $a$ parameter 1.8--90 times less than the Monte Carlo method.
The Hammersley set gives $a$ 3.5--95 times less than the Monte Carlo method.

Thus, in general, the benefit of polarization modeling when using Quasi-Monte Carlo is maintained.
We obtained the smallest gain in the case of a fractal cloud,
which has the most complicated geometry.

\subsection{Image-based results}\label{subsec:point_source_single_sc_results}

\begin{figure}
    \begin{minipage}[h]{0.495\linewidth}
        \includegraphics[width=\linewidth]{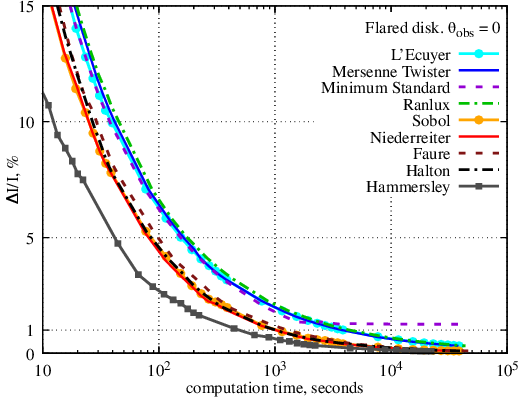}
    \end{minipage}
    \hfill
    \begin{minipage}[h]{0.495\linewidth}
        \includegraphics[width=\linewidth]{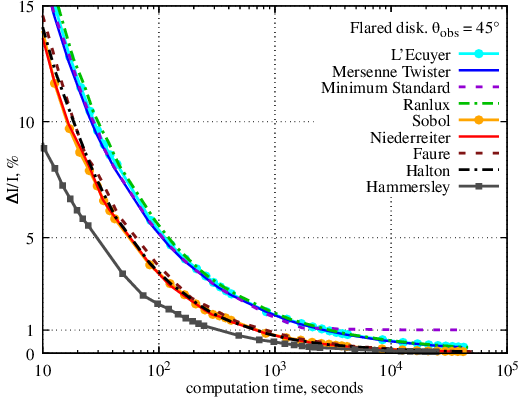}
    \end{minipage}
    \vfill
    \begin{minipage}[h]{0.495\linewidth}
        \includegraphics[width=\linewidth]{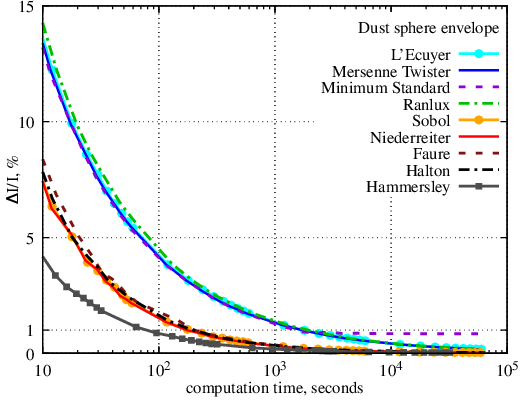}
    \end{minipage}
    \hfill
    \begin{minipage}[h]{0.495\linewidth}
        \includegraphics[width=\linewidth]{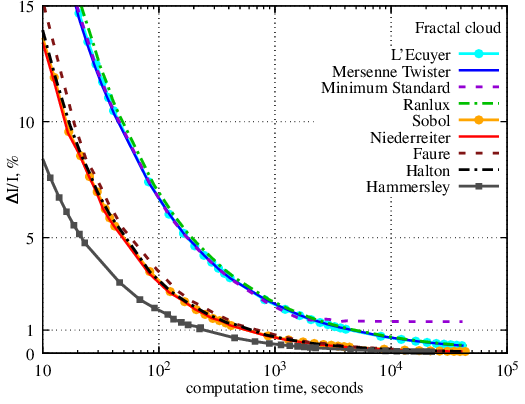}
    \end{minipage}
    \caption{The image-based difference norm for results obtained with different random number generators for
    the flared disk observed from the pole (top left panel),
    flared disk observed with $\theta_{obs} = 45$\textdegree\ (top right panel),
    the dust sphere envelope (bottom left panel), and the fractal dust cloud (bottom right panel).
    The star is considered a point source of radiation. Only single scatterings are simulated.}
    \label{fig:fractal_ps_1sc}
\end{figure}

The second way to compare the result is the image-based difference norm $\Delta I/I$~(\ref{norm}).
Its dependencies on time for the considered test geometries for all methods are presented in Fig.~\ref{fig:fractal_ps_1sc}.

Image-based difference norm $\Delta I/I$ is computed based on the pixel values.
For each pixel, we have random fluctuations like for $\delta I$ norm,
but accumulating the difference absolute values overall $200 \times 200$ pixels gives some kind of error averaging.
It leads to a significantly more stable function.
With our considered photon packet numbers, we even obtain a monotonous decrease in the $\Delta I/I$ norm for all geometries and generators.
It is possible to make simulations with close photon packet numbers and obtain fluctuations, but we do not see any sense in it for our study.

Again, the quality of the results produced by simulations with a Minimum Standard random number generator stops improving over time
after a certain point (see explanation in the previous subsection~\ref{subsec:point_source_single_sc_integral_results}).
Quasi-Monte Carlo demonstrates significantly better result quality compared to Monte Carlo.
The Quasi-Monte Carlo method with the Hammersley set outperforms other Quasi-Monte Carlo implementations.
A detailed analysis shows that there are two reasons for this.
Firstly, the solution with the Hummersley set gives a lower $\Delta I/I$ error than other Quasi-Monte Carlo implementations for the same number of photon packets.
Secondly, with an equal number of photon packets, calculations with the Hammersley set are noticeably faster (sometimes almost twice as fast).
It is most likely due to its photon packet modeling order, which gives more consistent access to density grid data and reduces the number of cache misses.

Now, on all panels in Fig.~\ref{fig:fractal_ps_1sc}, we can easily distinguish three families of curves:
Monte Carlo method (L'Ecuyer, Mersenne Twister, Minimum Standard, and Ranlux),
Quasi-Monte Carlo method (Sobol, Niederreiter, Faure, and Halton),
and Quasi-Monte Carlo method with the Hammersley set.
Inside every family, there are only slight differences between the curves:
Minimum Standard is suitable only for a small number of photon packets; Ranlux is a bit worse than L'Ecuyer or Mersenne Twister generators;
Faure is a bit worse compared to other Quasi-Monte Carlo generators (but much better than any Monte Carlo Generator).

The measured $\Delta I/I$ error depends not only on the error of the solution under study,
but also on the reference solution error.
We obtained the reference using the Monte Carlo, which has poor convergence.
For small computation times, the main factor is the error of the tested solution.
If the computation time is long, the reference error is also significant.
The reference error worsens the solution asymptotic behavior derived from the curve approximations.

Small fluctuations and close results of different generators inside one family allow us to produce a simple comparison.
We can fix the required $\Delta I/I$ error and check the computation time needed to achieve it.
For a brief result representation, we use a mean value for the Monte Carlo method and a mean value for the Quasi-Monte Carlo method again.
When the Minimum Standard solution does not achieve the required accuracy, we omit it from the averaging.
Table~\ref{table:ss_time} presents the computation time required to achieve different errors with three methods in all test geometries.

\begin{table}[htbp]
    \centering
    \begin{tabularx}{\textwidth}{>{\hsize=1.6\hsize} X | >{\hsize=0.8\hsize} X | >{\hsize=1\hsize}X  |>{\hsize=0.8\hsize}X  |>{\hsize=0.8\hsize}X }
                                        & $\Delta I/I$, \% & Monte Carlo & Quasi MC   & Hammersley \\ \hline
                                        &   $5$       & $1.6\cdot10^2$ & $89$           & $42$ \\
        Flared disk. $\theta= 0$        &   $1$       & $4.1\cdot10^3$ & $1.1\cdot10^3$ & $5\cdot10^2$ \\
                                        &   $0.3$     & $3.7\cdot10^4$ & $5.8\cdot10^3$ & $2.9\cdot10^3$ \\ \hline
                                        &   $5$       & $1.1\cdot10^2$ & $62$           & $31$ \\
        Flared disk. $\theta= 45$\textdegree & $1$    & $2.8\cdot10^3$ & $7.6\cdot10^2$ & $3.5\cdot10^2$ \\
                                        &   $0.3$     & $3.2\cdot10^4$ & $3.5\cdot10^3$ & $1.8\cdot10^3$ \\ \hline
                                        &   $5$       & $77$           & $19$           & $8$ \\
        Dust sphere envelope            &   $1$       & $1.7\cdot10^3$ & $1.9\cdot10^2$ & $80$ \\
                                        &   $0.3$     & $2\cdot10^4$   & $9.4\cdot10^2$ & $5\cdot10^2$ \\ \hline
                                        &   $5$       & $1.8\cdot10^2$ & $54$           & $22$ \\
        Fractal cloud                   &   $1$       & $4.8\cdot10^3$ & $6.6\cdot10^2$ & $2.5\cdot10^2$ \\
                                        &   $0.3$     & $3.9\cdot10^4$ & $3.0\cdot10^3$ & $1.5\cdot10^3$ \\
    \end{tabularx}
    \caption{Required computation time in seconds to achieve the specified $\Delta I/I$ error with different methods in single scattering simulations of the radiation from a point star.
        For each test geometry, the table shows the average values for the Monte Carlo method (L'Equier, Minimum Standard, Mersenne Twister, and Ranlux48 generators),
        the average value for the Quasi-Monte Carlo method (Sobol, Niederreiter, Faure, and Halton generators) and the value for the Quasi-Monte Carlo method with the Hammersley set.}
    \label{table:ss_time}
\end{table}

Table~\ref{table:ss_time} presents the same results as in Fig.~\ref{fig:fractal_ps_1sc} in a way that is more convenient for obtaining final results.
To achieve $5$\% $\Delta I/I$ error with the Quasi-Monte Carlo method, we need 1.8--4 less time than with the Monte Carlo method.
Quasi-Monte Carlo with the Hammersley set allows us to reach this accuracy 3.8--9 times faster than the Monte Carlo method.
For $1$\% $\Delta I/I$ error, Quasi-Monte Carlo and Quasi-Monte Carlo with the Hammersley set are 3.7--8.9 and 8--21 times faster than the Monte Carlo method.
Finally, for $0.3$\% $\Delta I/I$ error, they are 6--20 and 12--40 times faster, respectively.
We see significant computation speed up, which may be more than 10 times for some test geometries, errors, and methods.
Moreover, with the $\Delta I/I$ error decrease, the benefit from using the Quasi-Monte Carlo method increases.
Hence, the Quasi-Monte Carlo method has a better asymptotic behavior compared to the Monte Carlo method.

\medskip

In addition to the $\Delta I/I$ norm, we may introduce $\Delta Q/Q$ and $\Delta U/U$ norms for other Stokes parameters.
Depending on the chosen method and the geometry considered, the behavior of the polarization parameters $Q$ and $U$
differs slightly from the intensity $I$ behavior.
In all the test cases discussed above, we have $0.85 \Delta I/I < \Delta Q/Q < 1.1 \Delta I/I$.
When $U$ Stokes vector parameter differs from zero (for a disk observed with $\theta=45$\textdegree\ and a fractal dust cloud)
$0.95 \Delta I/I < \Delta U/U < 1.15 \Delta I/I$.
Thus, we see the image-based errors in different Stokes vector parameters are comparable.
It means that Quasi-Monte Carlo is good for polarization studies as well as for intensity studies.

\section{Multiple scattering case for a point source}\label{sec:point_sourse_ms}

For studying the methods performance in the multiple scatterings case, we consider the same four test geometries as in the previous section~(\ref{sec:point_source_single_sc}).
It makes the presentation of the results more understandable.
We simulate the first four scatterings for each photon packet.
Since a photon packet may leave the studied region before scattering all four times,
we obtain fewer photon packet scatterings for each subsequent scattering than for the previous one.

\begin{figure}
    \begin{minipage}[h]{0.495\linewidth}
        \includegraphics[width=\linewidth]{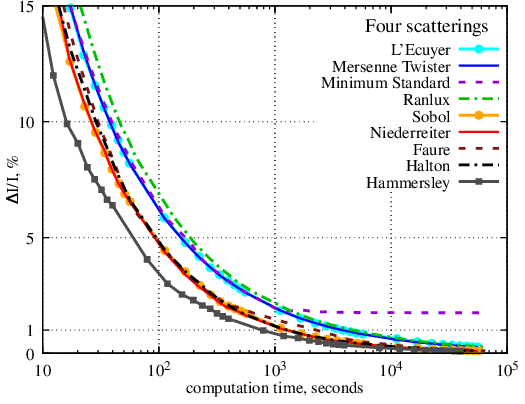}
    \end{minipage}
    \hfill
    \begin{minipage}[h]{0.495\linewidth}
        \includegraphics[width=\linewidth]{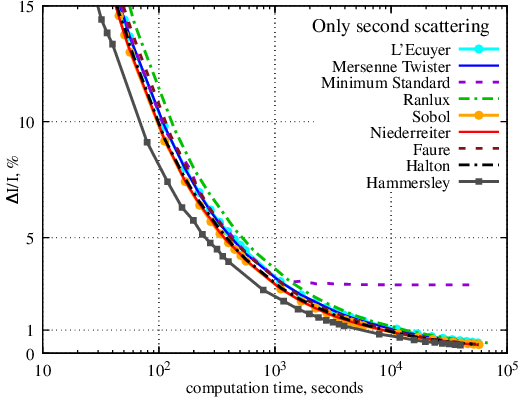}
    \end{minipage}
    \vfill
    \begin{minipage}[h]{0.495\linewidth}
        \includegraphics[width=\linewidth]{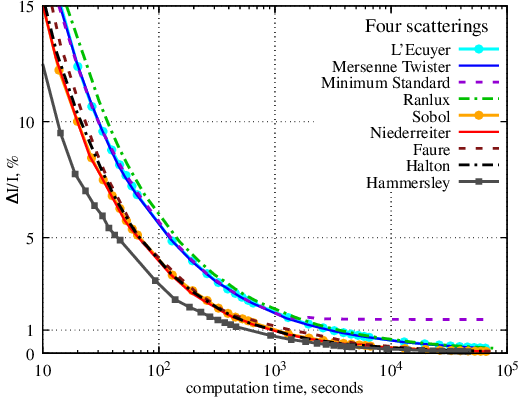}
    \end{minipage}
    \hfill
    \begin{minipage}[h]{0.495\linewidth}
        \includegraphics[width=\linewidth]{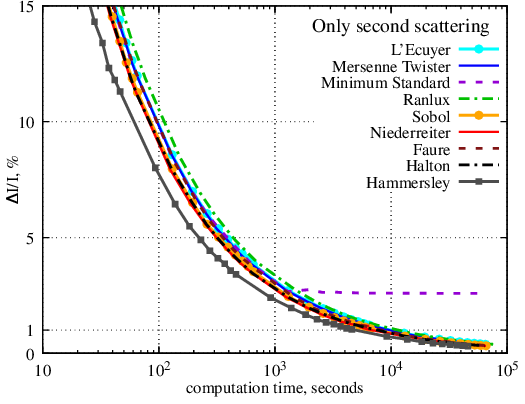}
    \end{minipage}
    \caption{Multiple scattering results for the flared disk observed from the pole (top panels)
        and the same flared disk observed with $\theta_{obs} = 45$\textdegree\ (bottom panels).
        The left panels demonstrate results for all four scatterings, while the right panels show only the second scatterings.
        The star is considered a point source of radiation.}
    \label{fig:ps_msc_disk}
\end{figure}

During the peel-off, we take into account the number of scatterings for a current photon packet.
Hence, we can compare the integral results and images for all the radiation coming to the observer (from all scatterings),
and separately only for single scatterings, only for double scatterings, and so on.

The quality of the result for single scattering in this chapter does not differ very much from that presented in the previous one:
we will get just minor fluctuations for the Monte Carlo method, as the pseudo-random numbers are used slightly differently.
In addition, in all cases, simulation time increases.
The photon packets are uniformly distributed.
Therefore, the simulation time increases proportionally for all considered photon packet numbers.
As a result, this increasement does not affect the relative comparison of the methods.

\begin{figure}
    \begin{minipage}[h]{0.495\linewidth}
        \includegraphics[width=\linewidth]{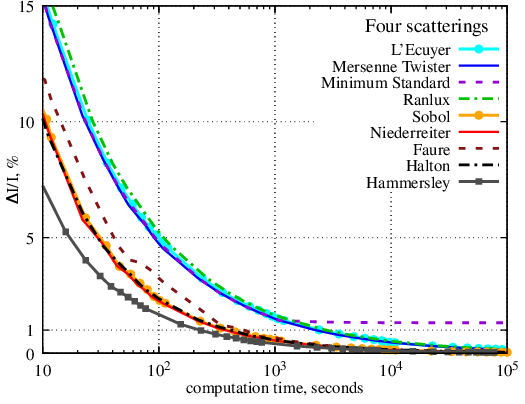}
    \end{minipage}
    \hfill
    \begin{minipage}[h]{0.495\linewidth}
        \includegraphics[width=\linewidth]{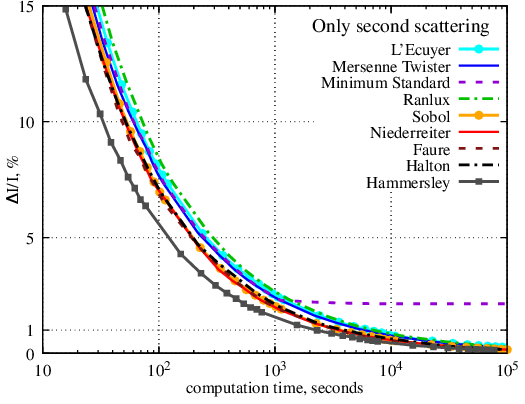}
    \end{minipage}
    \vfill
    \begin{minipage}[h]{0.495\linewidth}
        \includegraphics[width=\linewidth]{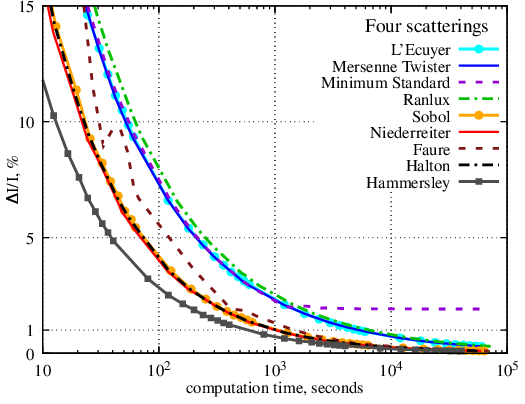}
    \end{minipage}
    \hfill
    \begin{minipage}[h]{0.495\linewidth}
        \includegraphics[width=\linewidth]{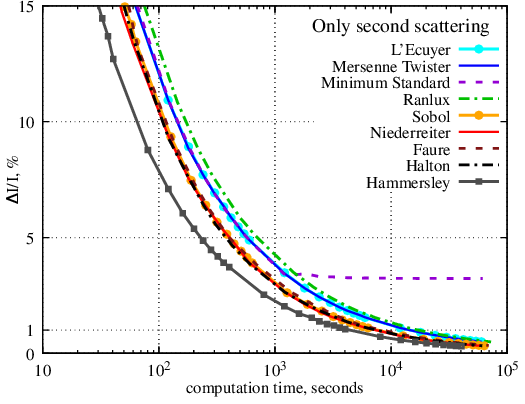}
    \end{minipage}
    \caption{Multiple scattering results for the dust sphere envelope (top panels) and the fractal dust cloud (bottom panels).
        The left panels demonstrate results for all four scatterings, while the right panels show only the second scatterings.
        The star is considered a point source of radiation.}
    \label{fig:ps_msc_sphere_and_fractal}
\end{figure}

Therefore, we can omit presenting the result for single scatterings here.
In the figures~\ref{fig:ps_msc_disk} and~\ref{fig:ps_msc_sphere_and_fractal} we present $\Delta I/I$ errors for all four simulated scatterings
and, separately, only for the second scattering.

The presented figures show that for the overall result, using Quasi-Monte Carlo gives a noticeable advantage over Monte Carlo.
When considering only the second scattering, the advantage of Quasi-Monte Carlo is clearly visible in some tests (the fractal dust cloud and the spherical envelope).
Although, it is less significant compared to the single scattering case.
In tests with the flared disk, the difference between Quasi-Monte Carlo and Monte Carlo is comparable to the spread of values in each of the generator groups.
The results of Quasi-Monte Carlo with the Hammersley set are better than the usual Quasi-Monte Carlo results in all tests for two reasons.
The Hammersley set gives a little better asymptotic convergence speed, but it is not very significant here.
The main effect is that for the same number of simulated photon packets, Quasi-Monte Carlo with the Hammersley set takes less computation time.

For the third and fourth scatterings, the difference between the methods becomes even smaller.
The $\Delta I/I$ error values are almost the same for all considered number generators for a fixed number of photon packets.
Only the Quasi-Monte Carlo method with the Hammersley set demonstrates better results due to the shorter computation time.

This effect has a simple explanation based on the error estimates from Section~\ref{subsec:error_estimation}.
For the Monte Carlo method, the asymptotic of the error does not depend on the problem dimension
unlike the Quasi-Monte Carlo method error.
Moreover, the higher the dimension of the problem, the later the error in the Quasi-Monte Carlo method ceases to decrease as $1/\sqrt{N}$ like in the Monte Carlo case.
When we take only single scatterings into account, a problem dimension is $3$.
For the second scatterings, it is already $6$, for the third scatterings~--- $9$, and for the fourth scatterings~--- $12$.
Consequently, the later scatterings we consider, the smaller the gain for them from using Quasi-Monte Carlo.

The contribution of subsequent scatterings to the final result decreases in our test problems.
The second scatterings make less contribution than the first scatterings and more than the third scatterings.
As a result, the significant gain in modeling the single scatterings is enough for a noticeable efficiency increase in the case of multiple scatterings,
although, for all scatterings except the first, we gain a tiny benefit.

We omit a table with exact time results for different $\Delta I/I$ errors and present only a short summary.
To achieve $5$\% $\Delta I/I$ error with the classical Quasi-Monte Carlo method and with the Quasi-Monte Carlo method with the Hammersley set,
we need 1.7--2.5 and 2.7--5.8 times less computation time than with the Monte Carlo method.
For $1$\% $\Delta I/I$ error, the Quasi-Monte Carlo method speed-ups are 2.6--4.8 and 5--9 times.
Finally, for $0.3$\% $\Delta I/I$ error, they are 5--8 and 8--13 times respectively.
The computations speed up is rather significant, though it is less than in the single scattering case.

\medskip

In the case of multiple scatterings, the analysis is more complicated than in the single scatterings case.
We can study Stokes vector parameters for different scatterings.
Given the large number of different geometries, we have many cases.
Mostly, the polarization parameters $Q$ and $U$ are close to zero for multiple scatterings.
In these cases, their relative errors rise significantly.
It is important to note that the $Q$ and $U$ accuracy does not depend on the selected method,
but depends on the geometry and scattering number in our tests.
When the value of the polarization parameter is non-zero, we usually get $\Delta U/U < 2 \Delta I/I$ and $\Delta Q/Q < 2 \Delta I/I$.
It means that using Quasi-Monte Carlo gives a gain either for all Stokes parameters or we get a result consistent with Monte Carlo for all of them.

\medskip

As an alternative to increasing the dimension in the quasi-random number generator for multiscattaring simulations,
one might consider leaving the generator dimension at 3.
Then, generated triples of quasi-random numbers are used to model successive scatterings,
regardless of what kind of scatterings they are.
This approach is not a correct use of quasi-random numbers and breaks the uniformity of their distribution.
When we try to apply this approach to simulate radiative transfer in practice, we get a terrible quality of the final result.
The resulting $\Delta I/I$ errors are huge. Moreover, we do not see any error decrease with increasing computation time.
We strongly discourage this approach.

\section{Single scattering case for a sphere source}\label{sec:single_scattering_sphere_source}

The consideration of single scatterings of photon packets from a sphere source of radiation instead of a point one
leads to an increase in the dimension of the problem by 2.
We have to simulate photon packets propagating in all directions from all points on the star's surface.
Of course, we do not consider photon packets going inside the star's surface, but this does not reduce the problem dimension.
So, we are going to calculate the integral in five-dimensional space.

While modeling radiation from a sphere source, we have two extreme cases:
\begin{enumerate}
    \item The source size significantly influences the result and is important for us.
    \item The size of the source is unimportant. Hence, we can consider it as point one.
\end{enumerate}
Actually, we cannot always understand without preliminary modeling that we have the second case.
In this regard, it would be fine if, in the second case, the modeling of radiation from a sphere source was as efficient
in terms of the ratio between the achieved accuracy to the calculation time as the modeling of radiation from a point source.

In this section, we consider four different circumstellar matter distributions.
In two distributions, the size of the star will be significant.
In the other two models, the star can be considered a point source of radiation.
So, we have three goals for this section:
\begin{enumerate}
    \item  compare methods for sphere sources of radiation;
    \item  make sure that the simulation results coincide with the simulation results for a point source when the star size is negligible;
    \item check how the increased dimension of the problem slows down the calculations.
\end{enumerate}
For the third part of the investigation, we consider the Quasi-Monte Carlo method with two approaches for emitting photon packets described in the section~\ref{subsec:photon_from_sphere_source}:
using the first and second dimensions of the fourth and fifth dimensions for the emission point on the star.

\subsection{Test cases}

We use two test cases from the section~\ref{sec:point_source_single_sc}: the flared disk observed with $\theta_{obs} = 45$\textdegree\ \ref{subsubsec:point_source_single_sc_tests_flared}
and the dust sphere envelope~\ref{subsubsec:point_source_single_sc_tests_sphere}.
For these two cases, we consider a sphere star with the radius $r_* = 2R_\odot$ now.

In these two tests, the star size is not really significant to us.
We can easily verify this: in both cases, we have a reference Monte Carlo solutions for a point source of radiation with $5 \cdot 10^{11}$ photon packets.
The Monte Carlo solution for a point source of radiation with $10^{10}$ photon packets has
the integral relative difference norm $\delta I \sim 10^{-5}$
and the image-based difference norm $\frac{\Delta I}{I} \approx 2.6\cdot10^{-3}$ in the flared disk case.
In the dust sphere envelope case, the norms are $\delta I \sim 10^{-5}$ and $\frac{\Delta I}{I} \approx 1.3\cdot10^{-3}$.

We simulated both models using the Monte Carlo method for a sphere star with $10^{11}$ photon packets and compared them with the reference solutions for a point source.
In the flared disk case, we received the norms are $\delta I \sim 10^{-5}$ and $\frac{\Delta I}{I} \approx 9\cdot10^{-4}$.
In the dust sphere envelope case, the norms are $\delta I \sim 10^{-5}$ and $\frac{\Delta I}{I} \approx 6\cdot10^{-4}$.

As we have seen above, the $\delta I$ norm is quite unstable.
We should not consider it to be a main one, but we should note that it is approximately the same in all cases.
It means that solutions for the point source and for a sphere source are close.
The $\frac{\Delta I}{I}$ is overall more stable.
Here, it demonstrates that, in both cases, the solution for a sphere source model with $10^{11}$ photon packets considered is closer to
the reference solution for a point source than the solutions for a point source with $10^{10}$ photon packets.
Moreover, the difference is about $2-3$ times, which is expected when we use the Monte Carlo method with 10-times more photon packets.

As a result, here we can actually consider the star as a point source of radiation.
These two test geometries are an extreme case for the sphere source radiation simulation.
In two more considered test geometries, the star radius should really matter.

It is worth noting that in reality, dust can not be very close to stars, where it heats up too much.
In this section, we completely ignore whether dust could exist so close to the star.

\subsubsection{Flared disk close to the star}\label{subsubsec:sphere_source_single_sc_tests_flared}

As a third test geometry, we use a flared disk that is close to the star.
The eq.~\ref{eq:flared_disk} describes the disk.
The sphere star radius $r_* = 2R_\odot$.
The flared disk model parameters are:
$R_i = 4 r_* = 8 R_\odot = 0.0372$ au and $R_d = 1$ au are inner and outer disk radii,
$h_0 = 0.004$ au is a disk scale height corresponding to a $R_0 = 0.04$ au radius,
$\alpha = 2$ is the radial density exponent, $\beta = 1$ is the flaring power,
and $\rho_0 = 2.4 \cdot 10^{-5}$ g cm$^{-3}$ is the density normalization constant.
This normalization constant gives the disk mass $\approx 0.09 M_\odot$.
The observer is positioned in direction with $\phi_{obs} = 0$ and $\theta_{obs} = 45$\textdegree.

The observed region radius is $1$~au.
A central circle with $r_*$ radius was masked out.

For this model, we prepared a reference solution for a spherical star, simulated $5 \cdot 10^{11}$ photon packets.
We also carried out simulations in this model with a point source of radiation, using $10^{11}$ photon packets.
The integral relative difference norm $\delta I \sim 0.14$
and the image-based difference norm $\frac{\Delta I}{I} \approx 0.14$.
Thus, taking the star size into account for this test geometry changes the result by 14\%.
It is a significant value that exceeds our typical errors considered in the paper.

\subsubsection{Dust sphere envelope close to the star}\label{subsubsec:sphere_source_single_sc_tests_sphere}

The fourth testing geometry is a dust sphere envelope close to the star.
Eq.~\ref{eq:sphere} describes the envelope.
The model parameters are:
$r_* = 5 R_\odot$ is the star radius,
$R_i = r_* = 5 R_\odot$ and $R_d = 2 r_* = 10 R_\odot$ are the inner and outer sphere radii,
and $\rho_0 = 6.29\cdot10^{-14}$ g cm$^{-3}$ is the density.
In this test, the envelope mass is $\approx 2.5\cdot10^{-11} M_\odot$.

The observer position direction is $\phi_{obs} = 0$ and $\theta_{obs} = 0$.
The observed region radius is $0.05$~au $(\approx 2.15 r_*)$.
A central circle with $r_*$ radius was masked out.

For this model, like the previous one, we prepared a reference solution for a spherical star, simulated $5 \cdot 10^{11}$ photon packets.
The simulations with a point source of radiation were done with $10^{11}$ photon packets.
The integral relative difference norm $\delta I \sim 0.07$
and the image-based difference norm $\frac{\Delta I}{I} \approx 0.07$.
Accounting for star size alters the result by 7\% in this test geometry.
It is less significant compared to the previous test geometry but also rather noticeable.

\subsection{Integral results}\label{subsec:sphere_source_single_sc_integral_results}

The integral norm behaves very unstable in these tests.
That is why it turns out to be a bad idea to compare all methods on its basis.
If we use $a \times t^{-0.5}$ approximation for comparison, the $a$ parameter also turns out to be unstable.
The error in $a$ determination is often higher than the difference between the $a$ values for different variations of Quasi-Monte Carlo.
We have Quasi-Monte Carlo with low-discrepancy sequences and the Hammersley set.
We tried the direct and inverse order of the photon packet generation.
Sometimes, these approaches have apparent differences in $a$, and sometimes not.

In general, we can conclude that the parameter $a$ in the $|\delta I|$ approximation is 3--10 times lower for Quasi-Monte Carlo than for Monte Carlo.
If the star can be considered a point source of radiation, the difference is about 10 times.
In this case, the inverse order of photon packet generation in Quasi-Monte Carlo reduces the error module by additional 2--3 times.
If the size of the star is significant, then the difference between the direct and inverse order of photon packet generation may be insignificant.
Using the Hammersley set usually results in a slight integral error reduction compared to other implementations of Quasi-Monte Carlo.

As in the case of a point source of radiation (Seq.~\ref{subsec:point_source_single_sc_integral_results}),
for the Stokes parameters $Q$ and $U$, we obtain a bit smaller difference in approximation parameter $a$ than for intensity $I$.
Nevertheless, there is a clear correlation between changes in the approximation parameter $a$ for different Stokes vector elements ($I$, $Q$, and $U$).
Using Quasi-Monte Carlo gives a comparable gain for all of them.
Using the Hammersley set and inverse order of photon generation increases this gain.

So, analysis of the integral results justifies using Quasi-Monte Carlo instead of Monte Carlo for a sphere star.

\subsection{Image-based results}\label{subsec:ss_results}

For clarity, we present the results for each test geometry in a separate figure.
On each figure, the left panel shows the results with the direct order of generation of the photon packets
(the first and second random numbers out of five are used to determine the emission location on the star's surface).
The right panels show the results for the inverse order of generation of photon packets
(the location of photon emission is determined by the fourth and fifth random numbers).
Because the order of random numbers is significant only for Quasi-Monte Carlo,
we show the same Monte Carlo solutions both in the left and right panels on all figures.

\begin{figure}
    \begin{minipage}[h]{0.495\linewidth}
        \includegraphics[width=\linewidth]{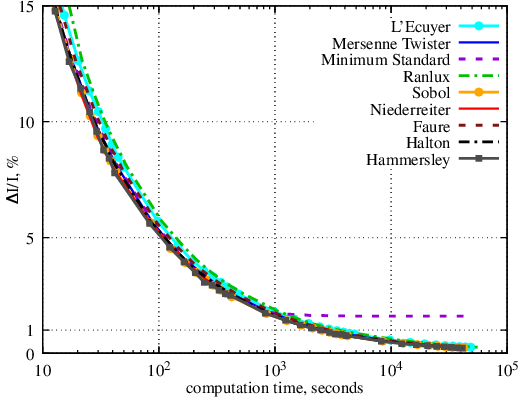}
    \end{minipage}
    \hfill
    \begin{minipage}[h]{0.495\linewidth}
        \includegraphics[width=\linewidth]{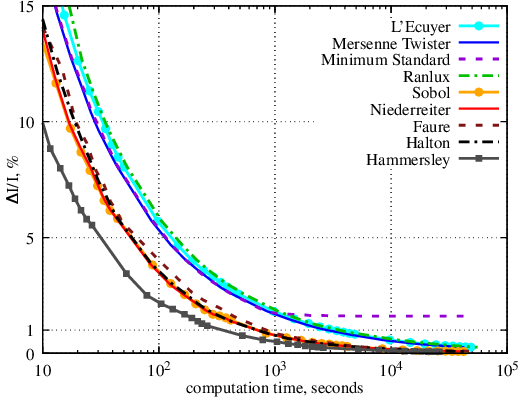}
    \end{minipage}
    \caption{The image-based difference norm for results obtained with different random number generators for the flared disk observed with $\theta_{obs} = 45$\textdegree.
        The star is considered a sphere source of radiation.
        Only single scatterings are simulated.
        The left panel shows the results for the direct order of photon packet generation, and the right panel~--- for the inverse order of photon packet generation. }
    \label{fig:ss_dsik_45}
\end{figure}

\begin{figure}
    \begin{minipage}[h]{0.495\linewidth}
        \includegraphics[width=\linewidth]{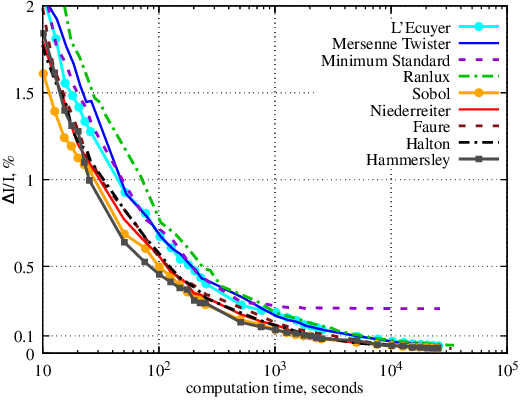}
    \end{minipage}
    \hfill
    \begin{minipage}[h]{0.495\linewidth}
        \includegraphics[width=\linewidth]{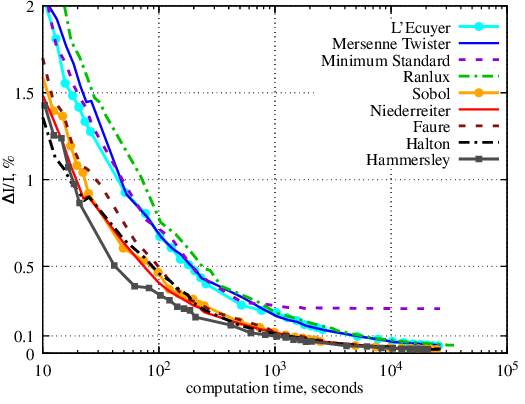}
    \end{minipage}
    \caption{The same as in Fig.~\ref{fig:ss_dsik_45} for the flared disk close to the star.}
    \label{fig:ss_dsik_45_close}
\end{figure}

\begin{figure}
    \begin{minipage}[h]{0.495\linewidth}
        \includegraphics[width=\linewidth]{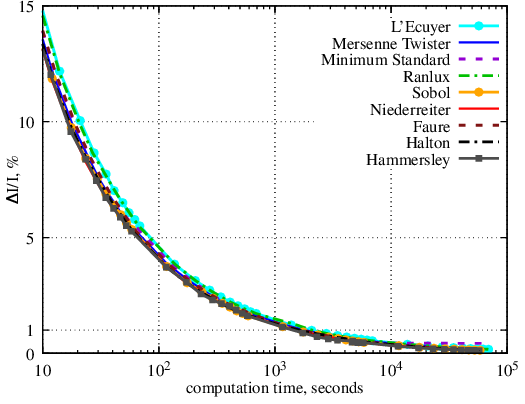}
    \end{minipage}
    \hfill
    \begin{minipage}[h]{0.495\linewidth}
        \includegraphics[width=\linewidth]{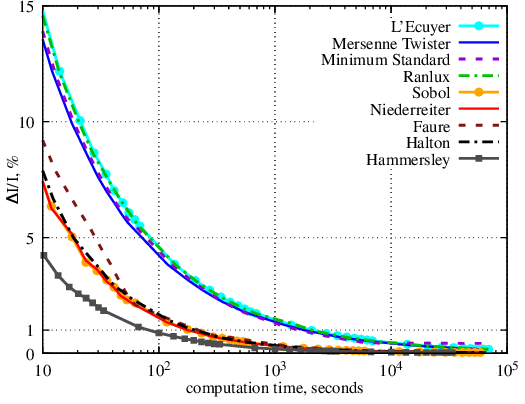}
    \end{minipage}
    \caption{The same as in Fig.~\ref{fig:ss_dsik_45} for the dust sphere envelope far from the star.}
    \label{fig:ss_sphere}
\end{figure}

\begin{figure}
    \begin{minipage}[h]{0.495\linewidth}
        \includegraphics[width=\linewidth]{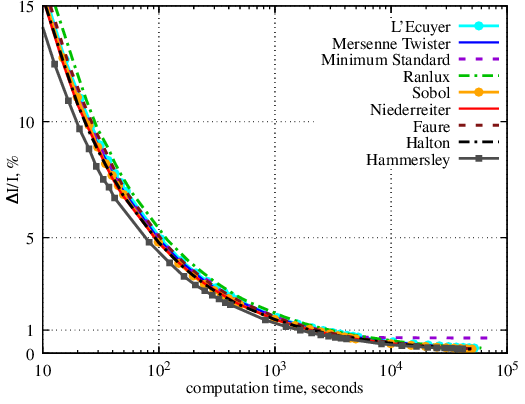}
    \end{minipage}
    \hfill
    \begin{minipage}[h]{0.495\linewidth}
        \includegraphics[width=\linewidth]{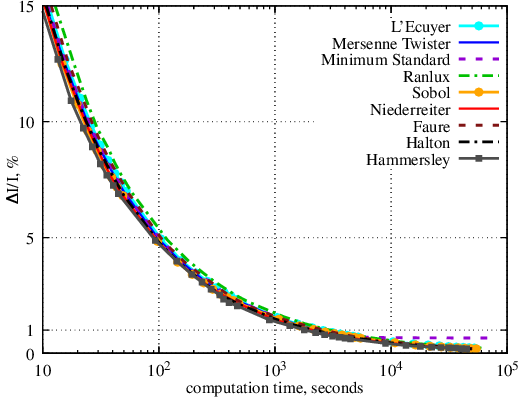}
    \end{minipage}
    \caption{The same as in Fig.~\ref{fig:ss_dsik_45} for the dust sphere envelope, which is close to the star.}
    \label{fig:ss_sphere_close}
\end{figure}

Figure~\ref{fig:ss_dsik_45} shows the $\Delta I/I$ errors for a flared disk located far from the star (described in the section~\ref{subsubsec:point_source_single_sc_tests_flared}).
Figure~\ref{fig:ss_dsik_45_close} demonstrates the results for a flared disk close to the star (from the model~\ref{subsubsec:sphere_source_single_sc_tests_flared}).
Figure~\ref{fig:ss_sphere} presents $\Delta I/I$ for a dust sphere envelope far from the star (described in the section~\ref{subsubsec:point_source_single_sc_tests_sphere}).
And finally, in Figure~\ref{fig:ss_sphere_close}~--- for a dust sphere envelope close to the star (from the model~\ref{subsubsec:sphere_source_single_sc_tests_sphere}).

A five-dimensional problem is more complicated than a three-dimensional one.
As a result, in some models, we see that $\Delta I/I$ error changes with time are not monotonous.
Nevertheless, its decrease with increasing computation time is clearly visible.

In almost all tests for the direct order of generation of photon packets,
we see only a minor difference between Monte Carlo and Quasi-Monte Carlo.
Only in the case of a flared disk close to the star (Fig.~\ref{fig:ss_dsik_45_close}) Quasi-Monte Carlo (and especially Quasi-Monte Carlo with the Hammersley set)
gives noticeable computation speed-up.

Increasing the problem dimension worsens the Quasi-Monte Carlo convergence.
We don’t always know whether we really need this increase to model a sphere radiation source.
That is why we proposed an approach with the reverse order of using random numbers for photon packet generation.
So the fourth and fifth dimensions with worse distribution are used for the emission position on the star.
When the star size is insignificant or is of little importance to us, this has a high impact,
as we move the main error to the least important part of the pipeline.
We clearly see this effect in three test cases out of four (the exception is the dust sphere envelope close to the star).

For greater clarity, we present the numerical results in the table~\ref{table:sphere_star_time}.
With direct order of photon generation, Quasi-Monte Carlo
(including the implementation with Hammersley set) turns out to be slightly faster than Monte Carlo.
Depending on the required $\Delta I/I$ error and test geometry, we get an acceleration up to 1.5--2 times.
We have never achieved 3 times acceleration in our tests.
It is much worse than the results for a point source of radiation.

Comparing the results for the models from the section~\ref{sec:point_source_single_sc},
we see that in these problems, modeling the radiation of a sphere source is a bit slower than the radiation of a point source.
Indeed, we use more random numbers, shift the photon emission point additionally,
and check during pilling-off that the photon packet does not intersect the star.
For the Monte Carlo method, it increases the computation time by about 10\%.
It is worth noting that we did not optimize this difference.
It can probably be reduced.
The computation time difference is approximately the same for Quasi-Monte Carlo with an inverse order of photon packet emission.
So, in the case of a point source of radiation, Quasi-Monte Carlo gives a significant gain compared to Monte Carlo,
regardless of whether we consider the star to be a point or not.

\begin{table}[htbp]
    \centering
    \begin{footnotesize}
    \begin{tabularx}{\textwidth}{>{\hsize=1.4\hsize} X | >{\hsize=0.6\hsize} X | >{\hsize=0.8\hsize}X  |>{\hsize=0.9\hsize}X  |>{\hsize=1.2\hsize}X|>{\hsize=1\hsize}X  |>{\hsize=1\hsize}X }
                        & $\Delta I/I$, \% & Monte Carlo & Quasi MC     & Hammersley     & Quasi~MC Inverse  & Hammersley Inverse\\ \hline
                        &   $5$       & $1.3\cdot10^2$ & $1.2\cdot10^2$ & $1.1\cdot10^2$ & $63$             & $33$ \\
        Flared disk     &   $1$       & $3.2\cdot10^3$ & $2.6\cdot10^3$ & $2.5\cdot10^3$ & $7.7\cdot10^2$   & $3.8\cdot10^2$ \\
                        &   $0.3$     & $3.7\cdot10^4$ & $2.1\cdot10^4$ & $2.4\cdot10^4$ & $3.5\cdot10^3$   & $2\cdot10^3$ \\ \hline
        Flared disk     &   $1$       & $54$           & $32$           & $25$           & $22$             & $18$           \\
        close to        &   $0.5$     & $2.1\cdot10^2$ & $1.2\cdot10^2$ & $85$           & $87$             & $42$ \\
        the star        &   $0.1$     & $5.3\cdot10^3$ & $1.9\cdot10^3$ & $2\cdot10^3$   & $1.1\cdot10^3$   & $9.7\cdot10^2$\\ \hline
        Dust sphere     &   $5$       & $83$           & $71$           & $68$           & $21$             & $8$\\
        envelope        &   $1$       & $2\cdot10^3$   & $1.6\cdot10^3$ & $1.5\cdot10^3$ & $2\cdot10^2$     & $84$\\
                        &   $0.3$     & $2.2\cdot10^4$ & $1.1\cdot10^4$ & $1.2\cdot10^4$ & $9.7\cdot10^2$   & $5.2\cdot10^2$ \\ \hline
        Dust sphere     &   $5$       & $104$          & $97$           & $78$           & $97$             & $90$\\
        envelope close  &   $1$       & $2.7\cdot10^3$ & $2.2\cdot10^3$ & $1.7\cdot10^3$ & $2.4\cdot10^3$   & $1.9\cdot10^3$\\
        to the star     &   $0.3$     & $4.2\cdot10^4$ & $2.2\cdot10^4$ & $2.1\cdot10^4$ & $2.4\cdot10^4$   & $2.6\cdot10^4$ \\
    \end{tabularx}
    \end{footnotesize}
    \caption{Computation time in seconds, which is required to achieve the specified $\Delta I/I$ error with different methods in single scattering simulations of the radiation from a sphere star.
        For each test geometry, the table shows the average values for the Monte Carlo method (L'Equier, Minimum Standard, Mersenne Twister, and Ranlux48 generators),
        the average value for the Quasi-Monte Carlo method (Sobol, Niederreiter, Faure, and Halton generators) and the value for the Quasi-Monte Carlo method with the Hammersley set.
        For the Quasi-Monte Carlo method, computation times for the implementation with the inverse photon packet generation approach are also presented.}
    \label{table:sphere_star_time}
\end{table}

In two cases (when the dust is far from the star), using the reverse order of generation of photon beams greatly accelerated the computations.
In the case of the flared disk close to the star, the speed up is approximately 1.5--2 times.
In the case of a dust sphere envelope close to the star, on the contrary, we received some deceleration.
We haven't explored where it came from.
Perhaps the order in which photon packets are generated is cache-friendly.
In general, we do not consider such a slowdown in the degenerate extreme case to be critical.

Thus, when we can consider the star to be a point source of radiation (or we are close to this situation),
Quasi-Monte Carlo gives noticeably better results compared to the Monte Carlo method.
If the size of the star is extremely important, Quasi-Monte Carlo gives an insignificant gain and still turns out to be better than Monte Carlo.

Additionally, it can be noted that in tests with high dimensions --- for a sphere star and for multiple scatterings,
the Quasi-Monte Carlo method with the Faure generator works worse than with the other generators.
The difference between the remaining generators does not seem significant.

\medskip

In two dust sphere envelope models, the $U$ Stokes parameter is zero.
Hence, we can examine $\Delta U/U$ norms only for the flared disk test from this section.
$\Delta Q/Q$ norm is suitable for all tests.
As a result we obtain  $0.73 \Delta I/I < \Delta Q/Q < 1.3 \Delta I/I$
and $0.86 \Delta I/I < \Delta U/U < 1.2 \Delta I/I$ for all considered approaches.
Thus, the image-based errors in $I$, $Q$, and $U$ Stokes vector parameters are comparable for a sphere star too.

\medskip

We omit a separate consideration of multiple scattering of radiation from a sphere source.
Multiple scattering always reduces the benefit of using Quasi-Monte Carlo.
The results for a sphere star do not yield anything fundamentally new.
Quasi-Monte Carlo still converges faster than regular Monte Carlo.
The gain depends on the number of simulated photon packets and test geometry (how significant the star size is).

\section{Conclusions}\label{sec:conclusion}

In our work, we considered using the Quasi-Monte Carlo method for modeling dust continuous radiative transfer instead of the traditional Monte Carlo method.
We investigated four different pseudorandom number generators for Monte Carlo: Minimum Standard \citep{Park1988}, \citet{LEcuyer1988}, Ranlux48 \citep{Luscher1994}, and Mersenne Twister \citep{Matsumoto2000}
For Quasi-Monte Carlo we used four low-discrepancy sequences: \citet{Halton1960}, \citet{Sobol1967}, \citet{Faure1981}, and \citet{Niederreiter1988}.
Additionally, we tested the Quasi-Monte Carlo method with the \citet{Hammersley1960} set.

We simulated single and multiple scatterings of radiation from both a point and a sphere source of radiation.
A detailed analysis was done for the light intensity.
$Q$ and $U$ Stokes vector parameters were also examined.
We found that errors of different Stokes vector parameters are correlated when using any method
(except for the cases of near-zero values of $Q$ and $U$).

Taking into account some errors in the computation time measurements and the result dependence on the test geometry
and some other factors (including random ones), our results can be summarized as follows:

\begin{enumerate}
\item When using the Monte Carlo method, there is no strong dependence on the choice of pseudorandom number generator.
    The main thing is that its period is long enough for the simulation task.
    In this regard, using the Minimum Standard generator is not recommended.

\item When using Quasi-Monte Carlo, we also get similar results when using different low-discrepancy sequences.
     When the problem dimension is high, the Faure sequence behaves worse than the others.
     There is no fundamental difference between the other sequences.

\item In all tests, Quasi-Monte Carlo performs better than Monte Carlo.
    The effect is noticeable both for integral values and for the image-based norm based on pixel-by-pixel comparison.

\item The computation time required to achieve a fixed image-based error using Quasi-Monte Carlo can be 10 times less than while using Monte Carlo.
    In some tests, we got gains up to 20 times.

\item In all tests except one, Quasi-Monte Carlo with the Hammersley set turns out to be faster than other implementations.
    The final speedup compared to Monte Carlo reaches 40 times in our tests.

\item The benefit from using Quasi-Monte Carlo decreases as the problem dimension becomes high.
    For a sphere source of radiation, the gain is less than for a point one.
    For multiple scatterings, it is less than for single scatterings.

\item It is possible to implement the simulation of a sphere star radiation in such a way that in cases where the star can actually be considered a point source,
    the solution with the Quasi-Monte Carlo method will not become worse due to the increased dimension of the problem.
\end{enumerate}

Using the Quasi-Monte Carlo method instead of Monte Carlo does not require major changes in the source code of the program: it is enough to replace the random number generator and add correct processing of the transition to the next photon packet simulation (see section~\ref{subsec:quasi_randoms_usage}).
It does not increase memory consumption.
Difficulties may arise only with multi-threaded implementation,
because it is impossible to use several fully independent random number generators with different seed values for different threads.
But this problem can also be solved.
Based on the above, we believe that replacing Monte Carlo with Quasi-Monte Carlo for dust continuous radiative transfer simulation is a good idea.
We recommend it as a logical development of existing approaches.

It is a rather challenging question whether significantly improved results can be achieved.
Our experiments from DGEM~\citep{Shulman2018} and new unpublished results show that for the single scatterings of radiation from a point source
one can obtain much better results using integration based on Riemann sums.
But we must remember two things.
First, even in the simplest case of single scatterings of radiation from a point source,
the effective implementation of Riemann sums integration for a radiative transfer is significantly more complex.
Secondly, the analysis of the error asymptotic behavior (see~\ref{subsec:error_estimation}) says that even for single scatterings from a sphere star
(and for multiple scatterings), we should not expect any gain from using integration with Riemann sums.
We tried it in practice.
Riemann sums integration use for a five-dimensional problem (single scatterings from a sphere star) turns out to be complex and ineffective.

It means that we can use Riemann sums only for single scatterings of radiation from a point source,
and then switch to the Monte Carlo (or maybe Quasi-Monte) method.
As an alternative, one can try to obtain a more efficient solution with Quasi-Monte Carlo.
The theoretical asymptotic behavior for Quasi-Monte Carlo assumes using infinite low-discrepancy sequences.
However, in practice, we often know in advance how many photon packets we are going to simulate.
It allows us to use approaches that take into account this quantity.
In our work, we used the Hammersley set. It almost always gave an advantage compared to the usual Quasi-Monte Carlo.
The math studies say that using Hammersley set in Quasi-Monte Carlo also gives a better theoretical asymptotic behavior.
There are more complex approaches that not only take into account the number of random realizations used,
but also involve a special selection of these values.
One of these realizations was proposed by~\citet{Owen1997}.
Such methods may speed up the radiative transfer simulations even more.

Of course, we expect that Quasi-Monte Carlo should be effective not only for the dust continuous radiation transfer,
but also for other radiative transfer simulations.
It opens up the possibility of improving computational approaches to many problems.

The test program is published under BSD-2-Clause license:
\url{https://github.com/sgshulman/DGEM}.


\def\acmtg{ACM Trans. Graph.}
\def\acmtms{ACM Trans. Math. Softw.}
\def\acmtmcs{ACM Trans. Model. Comput. Simul.}
\def\cacm{Commun. ACM}
\def\cmmph{Comput. Maths. Math. Phys.}
\def\bsmf{Bulletin de la Soci\'et\'e Math\'ematique de France}
\def\indmath{Indagationes Mathematicae}
\def\jnth{Journal of Number Theory}
\def\mcms{Monte Carlo Methods and Applications}
\def\nummath{Numerische Mathematik}
\def\siamjsc{SIAM Journal on Scientific Computing}
\def\paa{Proc. Akad. Amsterdam}
\def\procnyas{Proceedings of the New York Academy of Science}

\def\sovast{Sov. Astron.}
\def\araap{Annu. Rev. Astronomy and Astrophysics}
\def\apj{The Astrophysical Journal}
\def\apjss{Astrophysical Journal, Supplement Series}
\def\aac{Astronomy and Computing}
\def\aap{Astronomy and Astrophysics}
\def\basi{Bulletin of the Astronomical Society of India}
\def\cphc{Computer Physics Communications}
\def\lrcaph{Living Reviews in Computational Astrophysics}
\def\mnras{Monthly Notices of the Royal Astronomical Society}

\bibliography{ShulmanRefs.bib}

\end{document}